\PassOptionsToPackage{table}{xcolor}
\documentclass[sigconf]{acmart}

\usepackage{microtype}
\usepackage{graphicx}
\usepackage{subcaption}

\usepackage{multirow}

\usepackage{float}
\usepackage{booktabs}
\usepackage{placeins}

\usepackage{hyperref}

\usepackage[capitalize,noabbrev]{cleveref}

\AtBeginDocument{%
  }

\setcopyright{none}
\copyrightyear{}
\acmYear{}
\acmDOI{}
\acmConference[]{arXiv Preprint}{June, 2026}{}
\acmISBN{}

\begin{document}

\title{KernelSight-LM: A Kernel-Level LLM Inference Simulator}


\author{Xiteng Yao}
\authornote{This work was done while the first author was a summer intern at Amazon Web Services, under the mentorship and guidance of Taeho Kim and Hengzhi Pei, and others.}
\email{xtyao@bu.edu}
\affiliation{%
  \institution{Amazon Web Services}
  \city{Santa Clara}
  \state{CA}
  \country{USA}
}
\affiliation{%
  \institution{Boston University}
  \city{Boston}
  \state{MA}
  \country{USA}
}

\author{Taeho Kim}
\email{taehokim@amazon.com}
\affiliation{%
  \institution{Amazon Web Services}
  \city{Santa Clara}
  \state{CA}
  \country{USA}
}

\author{Hengzhi Pei}
\email{philepei@amazon.com}
\affiliation{%
  \institution{Amazon Web Services}
  \city{Santa Clara}
  \state{CA}
  \country{USA}
}

\author{Xinle Liu}
\email{sliuxl@amazon.com}
\affiliation{%
  \institution{Amazon Web Services}
  \city{Santa Clara}
  \state{CA}
  \country{USA}
}

\author{Kyle Ulrich}
\email{ulrichkr@amazon.com}
\affiliation{%
  \institution{Amazon Web Services}
  \city{Santa Clara}
  \state{CA}
  \country{USA}
}

\author{Leonard Lausen}
\email{lausen@amazon.com}
\affiliation{%
  \institution{Amazon Web Services}
  \city{Santa Clara}
  \state{CA}
  \country{USA}
}

\author{Ashish Khetan}
\email{khetan@amazon.com}
\affiliation{%
  \institution{Amazon Web Services}
  \city{Santa Clara}
  \state{CA}
  \country{USA}
}

\author{Xiang Song}
\affiliation{%
  \institution{Amazon Web Services}
  \city{Santa Clara}
  \state{CA}
  \country{USA}
}

\author{George Karypis}
\affiliation{%
  \institution{Amazon Web Services}
  \city{Santa Clara}
  \state{CA}
  \country{USA}
}

\author{Martin Herbordt}
\email{herbordt@bu.edu}
\affiliation{%
  \institution{Boston University}
  \city{Boston}
  \state{MA}
  \country{USA}
}

\renewcommand{\shortauthors}{Xiteng et al.}

\begin{abstract}
As large language models (LLMs) move into production serving, practitioners must rapidly
evaluate inference performance across diverse hardware, models, and serving parameters to meet cost and latency targets. However, the end-to-end behavior of LLMs couples serving-layer policies with low-level GPU kernel execution and rapidly evolving architectures, forcing slow, deployment-specific benchmarking that is hard to generalize.

We present \textsc{KernelSight-LM}, a fine-grained inference simulator that models token-level execution and produces kernel-level latency breakdowns. It decomposes each serving step into a roofline kernel model with a learned efficiency term, a communication model, and a host-overhead model, composed through a discrete-event scheduler that also captures sophisticated mechanisms including prefix caching and continuous batching.

KernelSight-LM offers two prediction tiers that trade target-GPU data for accuracy. The
\emph{cross-generation} tier uses no target-GPU measurements, only hardware specifications and kernel microbenchmarks from previously profiled GPUs, and predicts per-kernel latency on an unseen GPU generation to $12.1\%$ error, a $1.8\times$ improvement over the roofline baseline ($22.0\%$). 

A second \emph{target-measured} tier adds one model-agnostic kernel-microbenchmark sweep on the target GPU, sharpening per-kernel error to $3.8\%$, a $7.3\times$ improvement over a comparable baseline ($27.7\%$). Both tiers require far less target-GPU data than the prior systems they extend.

Composed through our simulator, these kernel predictions yield end-to-end median (p50) errors across six model families of $15.4\%$, $12.8\%$, and $3.0\%$ (TTFT, TPOT, throughput) in the cross-generation tier and $14.3\%$, $6.2\%$, and $2.7\%$ in the target-measured tier, matching dedicated profiling tools while collecting far less on-device data. Beyond prediction, its kernel-level bottleneck breakdowns support hardware/software co-design and capacity planning.

\end{abstract}

\ccsdesc[500]{Computing methodologies~Modeling and simulation}
\ccsdesc[500]{Computing methodologies~Machine learning}
\ccsdesc[500]{Computer systems organization~Parallel architectures}

\keywords{Large language models, LLM inference serving, performance prediction, GPU kernel latency, roofline model, cross-generation prediction, discrete-event simulation}

\maketitle

\section{Introduction}

Large language models (LLMs)~\cite{vaswani_attention_2023, brown_language_2020} are
deployed at scale for interactive assistants~\cite{mahmood_llmpowered_2023},
retrieval-augmented generation~\cite{lewis_retrievalaugmented_2020}, and beyond,
where inference dominates operational cost and determines user experience through
latency and tail-latency service-level objectives (SLOs)~\cite{zhong_distserve_2024,
hong_sola_2025}. Meeting these SLOs is hard: production serving runs under
heterogeneous request mixes, bursty arrivals, and high concurrency, so serving
metrics are difficult to predict ahead of deployment.

Teams therefore rely on deployment-specific benchmarking and iterative
experimentation to choose model variants, configurations, and hardware. This is slow,
expensive, hard to reproduce, and—critically—unavailable for early capacity planning
or procurement on hardware not yet owned.

Prior work reduces this reliance along three directions, each confined to one regime.
NeuSight~\cite{lee_forecasting_2025} forecasts performance on unseen GPUs but exposes
no serving-policy or tail-latency effects. AIConfigurator~\cite{xu_aiconfigurator_2026} interpolates
measured per-device operator tables but cannot extrapolate to an unmeasured GPU and
approximates the small kernels. Vidur~\cite{agrawal_vidur_2024} simulates end-to-end
serving faithfully but ties execution timing to per-(model,\,GPU) profiling.

We present \textsc{KernelSight-LM}, which integrates and extends Vidur with
kernel-level latency prediction, organized as \emph{tiers} by available target data:
datasheet-only zero-shot (Tier~A, the \emph{cross-generation} tier) and target microbenchmarks (Tier~B, the \emph{target-measured} tier). A single model thus spans profiling-free extrapolation to high-fidelity in-distribution prediction, while exposing
serving-policy interactions and system overheads for bottleneck diagnosis and what-if
analysis.

In summary, our contributions are:
\begin{itemize}
\item \textbf{Zero-shot cross-generation prediction (Tier~A).} We design a
roofline$\times$efficiency model with learned per-family heads that predicts per-kernel
latency on an unseen GPU generation from datasheet specifications alone, with no target
measurements (\S\ref{sec:tierA}).

\item \textbf{Target-measured refinement (Tier~B).} We extend the same model with a
one-shot, model-agnostic microbenchmark sweep that interpolates a measured efficiency
grid to sharpen in-distribution accuracy (\S\ref{sec:tierB}).

\item \textbf{Serving-aware simulation.} We re-implement vLLM~v1's scheduler
step-for-step and add a two-tier (GPU/disk) prefix cache and explicit chunked-prefill
pricing, so batching and caching match the real engine's per-step batch composition
(\S\ref{sec:sim}).

\item \textbf{Modeled host calibration.} We measure host-side overhead terms—CPU/GPU
overlap, scheduling tail, admission, launch, and TP synchronization—per
(host,\,GPU,\,engine) for more accurate Tier~B latency prediction (\S\ref{sec:assembly}).
\end{itemize}



\section{Background}

\subsection{Modern LLMs and Inference Frameworks}
Modern production LLMs are predominantly decoder-only Transformer models served in an autoregressive setting. Inference naturally decomposes into two distinct phases: prefill, which processes the input prompt in parallel, and decode, which generates output tokens sequentially~\cite{zhong_distserve_2024, patel_splitwise_2024}. This decomposition is fundamental to performance characterization: prefill exhibits substantial parallelism and is typically compute-bound, while decode introduces sequential dependencies and is often memory-bound due to per-token attention over growing KV caches.

LLM inference frameworks, therefore, optimize along two axes: (i) efficient operator execution and (ii) serving-level techniques that balance throughput and latency. Systems such as vLLM~\cite{kwon_efficient_2023a} popularize KV-cache-aware memory management and continuous batching~\cite{_orca_} to sustain high throughput under variable-length requests, while vendor stacks such as TensorRT-LLM~\cite{nvidiacorporation_tensorrtllm_2023} emphasize kernel fusion, quantization, and hardware-specific tuning to maximize performance on particular GPU families.

In production, service-level objectives (SLOs) are the most important metrics used to show LLM serving quality. Several key SLOs are (a) time-to-first-token (TTFT), the delay from request submission to the first generated token, (b) time-per-output-token (TPOT), the average inter-token time during decoding after the first token, and (c) output throughput, commonly measured in generated tokens per second (tok/s) under a specified workload and concurrency. Measuring and optimizing these metrics is central to maintaining serving quality in deployed LLM systems.

\begin{figure}[t]\centering
  \includegraphics[width=\columnwidth]{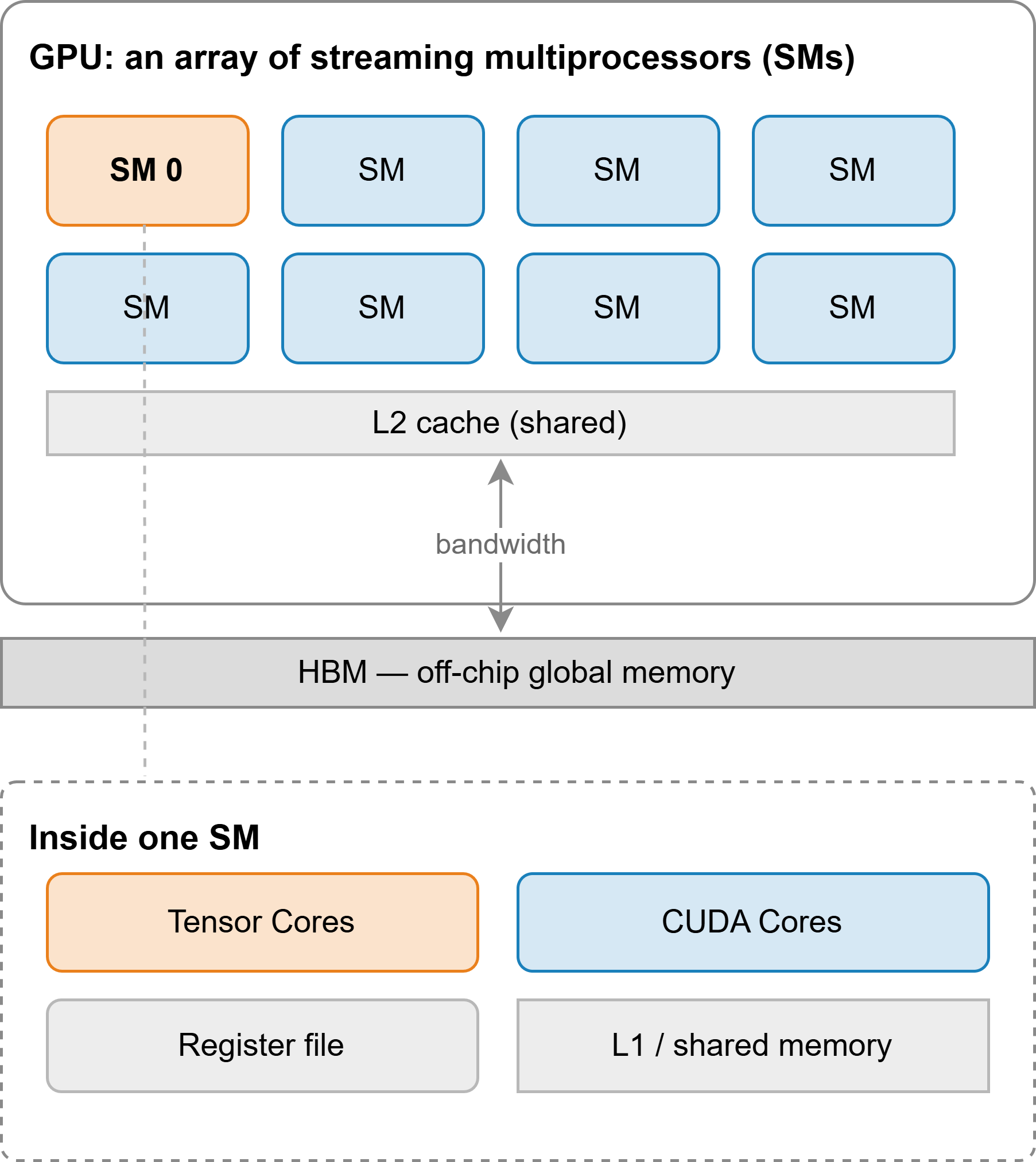}
  \caption{A GPU is an array of streaming multiprocessors (SMs) sharing an L2 cache and
  off-chip HBM; each SM holds tensor cores, CUDA cores, a register file, and L1/shared
  memory.}
  \label{fig:gpu_arch}
\end{figure}

\subsection{Graphics Processing Units}
\label{sec:gpu}
A GPU is a massively parallel processor built from an array of streaming
multiprocessors (SMs) over a memory hierarchy of per-SM registers and L1/shared
memory, a shared L2 cache, and off-chip high-bandwidth memory (HBM), as shown in
Figure~\ref{fig:gpu_arch}. Each SM runs thousands of threads concurrently and contains
both general-purpose CUDA cores and specialized tensor cores; the tensor cores
accelerate the dense matrix multiplications that dominate LLM transformer layers at far
higher throughput than CUDA cores~\cite{nvidiacorporation_nvidia_2024}.

\begin{figure}[t]\centering
  \includegraphics[width=\columnwidth]{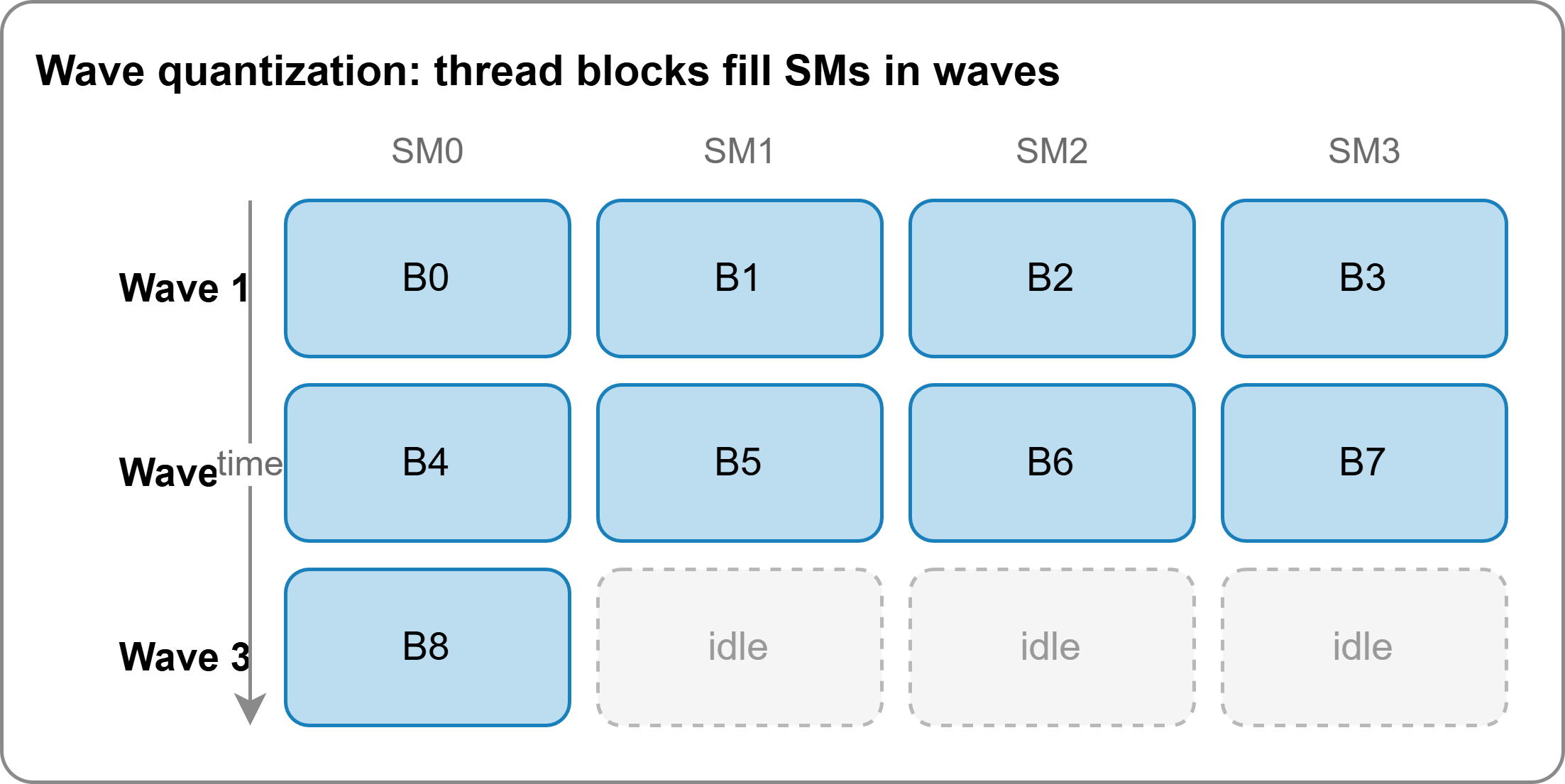}
  \caption{Wave quantization. A kernel's $B$ thread blocks are scheduled onto the
  $N_{\mathrm{SM}}$ SMs in \emph{waves}; when $B$ is not a multiple of $N_{\mathrm{SM}}$
  the final wave is partial and leaves SMs idle (here $9$ blocks on $4$ SMs span $3$
  waves, leaving $3$ of $4$ SMs idle in the last), inflating execution time by
  $u=\lceil B/N_{\mathrm{SM}}\rceil\,N_{\mathrm{SM}}/B \ge 1$ in the compute roofline
  (Eq.~\eqref{eq:master}).}
  \label{fig:wave_quant}
\end{figure}

Two hardware limits bound a kernel's performance: peak compute throughput (FLOP/s), set
largely by the tensor cores, and memory bandwidth (GB/s) to HBM. Which one binds depends
on the kernel's arithmetic intensity relative to their ratio (\S\ref{sec:roofline}). A
second, finer effect is how a kernel's thread blocks tile onto the $N_{\mathrm{SM}}$
SMs: when the block count is not a multiple of $N_{\mathrm{SM}}$, the final scheduling
\emph{wave} is partial and leaves SMs idle (Figure~\ref{fig:wave_quant}), a
quantization our kernel model prices explicitly (\S\ref{sec:eff-pred}).

\subsection{Roofline Model and Arithmetic Intensity}
\label{sec:roofline}
The roofline model~\cite{williams_roofline_2009} bounds an operation's achievable
throughput by its arithmetic intensity (AI), the amount of computation performed per
byte of data movement:

\begin{equation}
\mathrm{AI} = \frac{\text{arithmetic operations (FLOPs)}}{\text{data movement (bytes)}}.
\end{equation}

\begin{figure}
  \centering
  \includegraphics[width=0.92\columnwidth]{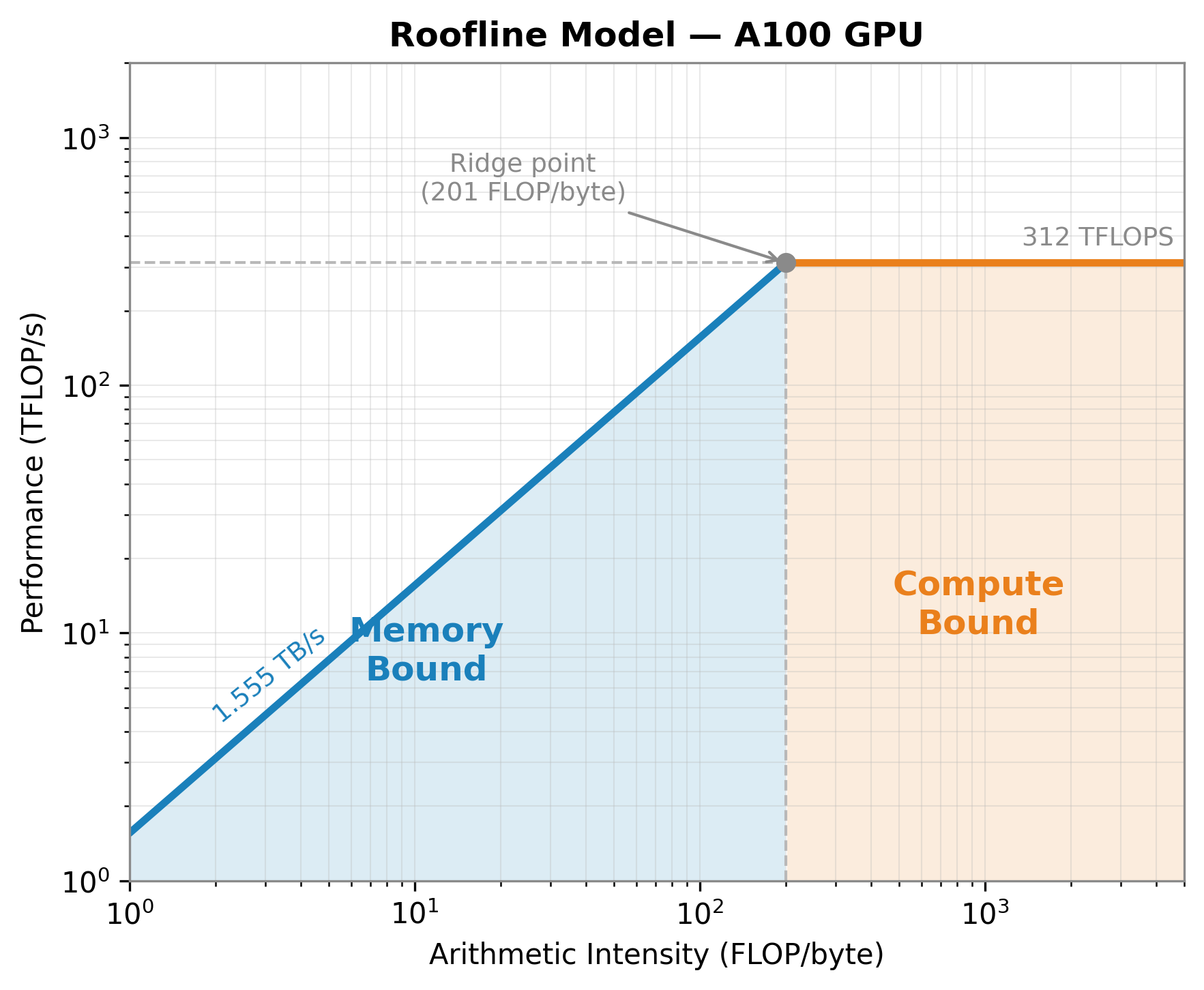}  
  \caption{An example of a roofline plot of A100 GPU.}
  \label{fig:roofline_basic}
\end{figure}

A roofline plot places an operation at its AI on the $x$-axis (FLOP/byte) and its
achieved throughput on the $y$-axis (FLOP/s). As shown in
Figure~\ref{fig:roofline_basic}, two ceilings bound performance: a horizontal
\emph{compute ceiling} at the peak compute throughput, and a sloped \emph{memory
ceiling} equal to peak bandwidth $\times$ AI. They meet at the \emph{ridge point}:
operations with AI below it are memory-bound (limited by the sloped ceiling), and
those above it are compute-bound (limited by the flat ceiling).

Because each GPU has its own peak compute and bandwidth, its ceilings and ridge point
sit at different positions. The same operation can therefore be memory-bound on one
GPU and compute-bound on another, which is part of what makes performance hard to
predict across GPU generations.



\begin{table*}[t]
\centering
\caption{KernelSight-LM's Tier~A and Tier~B as models, against the strongest prior
system in each regime. Tier~A is the cross-generation counterpart to NeuSight,
Tier~B the in-distribution counterpart to AIConfigurator; both are kernel-level and
plug into a full serving simulator (cf.\ Vidur). \checkmark~full, $\sim$~partial,
--~absent.}
\label{tab:tier-comparison}
\small
\begin{tabular}{l c c c c c}
\toprule
 & & & & \multicolumn{2}{c}{\textbf{KernelSight-LM (ours)}} \\
\cmidrule(lr){5-6}
\textbf{Property} & \textbf{NeuSight} & \textbf{AIConfig.} & \textbf{Vidur} & \textbf{Tier~A} & \textbf{Tier~B} \\
\midrule
Predict unbenchmarked GPU (cross-gen) & \checkmark$^{1}$ & -- & -- & \checkmark & -- \\
In-distribution from target measurements & $\sim$ & \checkmark & \checkmark & -- & \checkmark \\
Kernel-level granularity & -- & -- & -- & \checkmark & \checkmark \\
No per-(model\,$\times$\,GPU) profiling & \checkmark & $\sim$ & -- & \checkmark & \checkmark \\
Models all kernels $+$ serving-matched attention & -- & $\sim^{2}$ & -- & \checkmark & \checkmark \\
Full serving simulation (batching, chunked prefill, prefix cache) & -- & $\sim^{3}$ & \checkmark & \checkmark & \checkmark \\
Host/CPU overhead channel & -- & -- & \checkmark & \checkmark & \checkmark \\
End-to-end metrics (TTFT/TPOT/throughput) & $\sim$ & \checkmark & \checkmark & \checkmark & \checkmark \\
\bottomrule
\end{tabular}

\vspace{2pt}
{\footnotesize
$^{1}$Single-generation extrapolation (Ampere$\to$Hopper); no Blackwell.\quad
$^{2}$Measures GEMM/attention; approximates elementwise/memory kernels analytically.\quad
$^{3}$Analytical assembly, not event-level scheduling.}
\end{table*}

\section{Related Work}

Tools for estimating LLM inference performance fall into two broad families: those that
\emph{measure} execution latencies on the target hardware (accurate in-distribution,
but tied to GPUs they have already profiled) and those that \emph{forecast} latencies
on GPUs not yet benchmarked (portable across hardware, but blind to serving dynamics).
We summarize each family with a representative system below and identify the gap our
work closes: no prior system does both. Other tools target adjacent problems—e.g.,
interconnect/collective simulation and token-level modeling~\cite{won_astrasim20_2023,
rashidi_astrasim_2020, wu_tokensim_2025, lin_apex_2025, cho_llmservingsim20_2025}—and
are out of scope, as they do not jointly address cross-device extrapolation and
serving-policy interactions.

\subsection{Measurement-Based Estimation}
Both discrete-event simulators and table-interpolation tools estimate serving
performance from execution latencies \emph{measured on the target hardware}, so a new
GPU or model variant requires a fresh profiling run, and neither extrapolates to an
unbenchmarked device. They differ, however, in whether they model serving dynamics.
Discrete-event simulators such as Vidur~\cite{agrawal_vidur_2024} reproduce the serving
pipeline event-by-event—batching, scheduling, queueing, and prefill/decode
interaction—and report end-to-end throughput, latency, and tail metrics, which makes
them well-suited to capacity planning and policy exploration. Table-interpolation tools
such as AIConfigurator~\cite{xu_aiconfigurator_2026} instead interpolate per-(device,
framework) operator measurements and assemble end-to-end metrics \emph{analytically},
performing no event-level scheduling, to recommend deployment configurations; this is
fast and accurate in-distribution, but it models GPU execution only (no host/CPU
channel) and approximates the small elementwise and memory kernels with a fixed
analytical form.

\subsection{Cross-Device Performance Forecasting}
A second approach predicts operator latency on GPUs not yet benchmarked~\cite{yu_habitat_2021, zhang_nnmeter_2021, qi_paleo_2017, imai_predicting_2024}. A
representative system, NeuSight~\cite{lee_forecasting_2025}, symbolically traces a
model with PyTorch FX~\cite{reed_torch_2022} into primitive operators (linear layers,
batched matrix multiplications, vector operations) and predicts each with a learned
wave-based model—bandwidth-utilization factors parameterized by GPU architectural
features such as SM count, memory bandwidth, L2 size, and peak FLOPs—then aggregates
per-operator latencies into an end-to-end estimate and projects across GPU
generations. Such forecasters, however, do not model production serving behavior—request
arrivals, queueing, batching/scheduling, KV-cache management~\cite{lee_infinigen_2024},
and prefill--decode interaction~\cite{zhong_distserve_2024}—that drives SLO decisions.

\subsection{Gaps in Existing Approaches}
Each family is confined to one regime, leaving a gap between serving fidelity and
hardware portability.

\paragraph{Measurement-based estimation cannot extrapolate.}
Both simulators and table-interpolation tools depend on latencies measured per
model--hardware pair, so reaching a new GPU or model variant requires a fresh profiling
run—time-consuming or impossible without the target accelerator.

\paragraph{Cross-device forecasting omits serving dynamics.}
Operator-level forecasters transfer latency across GPUs but do not model request
arrivals, queueing, batching/scheduling, KV-cache management, or prefill--decode
interaction—central to SLO-driven decisions.

\paragraph{Host-side overheads are under-modeled.}
Across both families, host-side overheads—CPU/GPU overlap, scheduling and launch costs,
and tensor-parallel synchronization—are absent or only implicitly captured in profiled
latency tables.

\paragraph{KernelSight-LM spans both axes.}
We combine cross-generation extrapolation with full serving-dynamics simulation and an
explicit host-overhead channel, all at kernel granularity. Tier~A predicts an
unbenchmarked GPU from its datasheet alone (the regime of NeuSight), while Tier~B
sharpens accuracy from a single target microbenchmark sweep (the regime of
AIConfigurator)—both feeding the same discrete-event serving simulator (cf.\ Vidur).
Table~\ref{tab:tier-comparison} summarizes these capabilities against the strongest
prior system in each regime.

\begin{figure*}[ht]
  \centering
  \includegraphics[width=\textwidth]{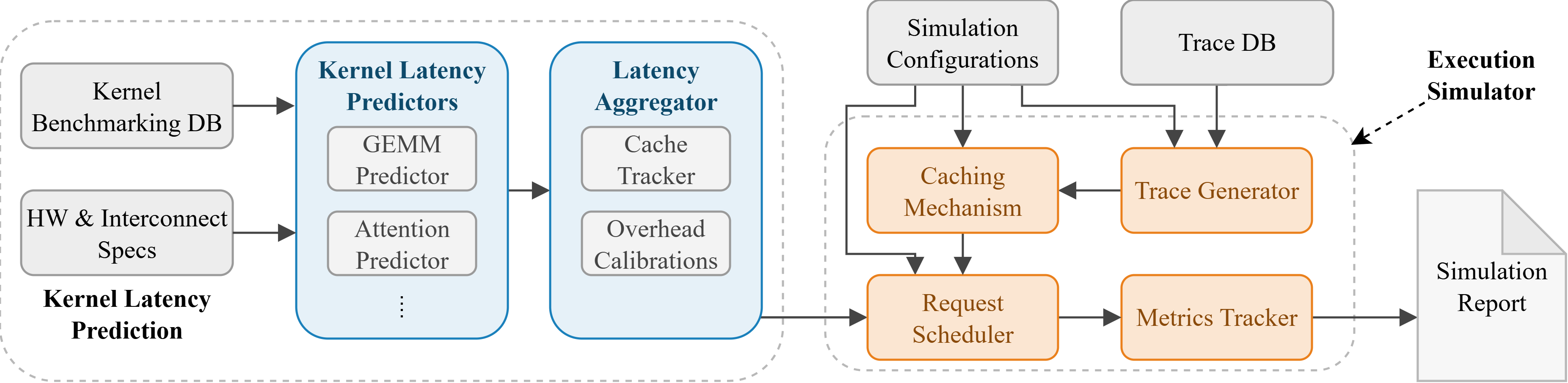}  
  \caption{An overview of KernelSight-LM's architecture. The left-hand input boxes are the data the user provides (hardware and interconnect specifications) together with the pre-collected kernel-benchmark database. Blue boxes are the kernel-latency predictors and latency aggregator (Tier~A and Tier~B); orange boxes are the discrete-event simulator runtime.}
  \label{fig:overview}
\end{figure*}

\section{Tool Architecture and Methodologies}

\subsection{System Overview}

As shown in Figure~\ref{fig:overview}, our KernelSight-LM consists of two major components. The left half depicts our kernel-level latency predictors. The kernel predictor operates in two tiers, Tier~A and Tier~B. Both modes predict kernel latencies using workload characteristics, target GPU specifications, and a pre-collected, model-agnostic microbenchmarking database. However, Tier~B has access to the on-target microbenchmarking dataset, giving it higher accuracy.
The right half shows the simulator runtime, built upon Vidur's execution simulator~\cite{agrawal_vidur_2024}.

End-to-end, a single batch flows as follows: the scheduler emits a batch descriptor (decode tokens, prefill-chunk lengths, per-request KV lengths); the predictor maps each dispatched kernel to a latency via the roofline$\times$efficiency model of Eq.~\eqref{eq:master}, sums these with the tensor-parallel all-reduce into a step time (Eq.~\eqref{eq:assembly}), and returns that duration to the discrete-event scheduler, which advances the simulation clock and accumulates the per-request TTFT, TPOT, and throughput it reports.

\subsection{Execution Simulator}

\label{sec:sim}
Our simulator backend follows the same mechanism as Vidur. It operates as a discrete-event simulator rather than executing actual GPU workloads. Requests arrive according to a synthetic generator or replay trace, and a global scheduler dispatches them to replicas based on configurable policies. Each replica maintains its own scheduling state and forms batches according to the selected algorithm. When a batch is ready for execution, the simulator invokes our kernel-level execution time predictor: given the batch composition, the predictor aggregates individual kernel latencies into a total batch execution time. This predicted duration is then used to schedule an event to advance the simulation clock by the predicted execution time. This architecture makes it possible for us to switch the latency prediction mechanisms while using the sophisticated request scheduling and simulation workflow.

Our extension to Vidur also inherits Vidur's rich features for simulating different inference scenarios, such as pipeline and data parallelism. Beyond that, our execution simulator enables cross-generation inference prediction and more faithful simulation of continuous batching, chunked prefill, and related mechanisms.

\begin{figure*}[t]
  \centering
  \includegraphics[width=\textwidth]{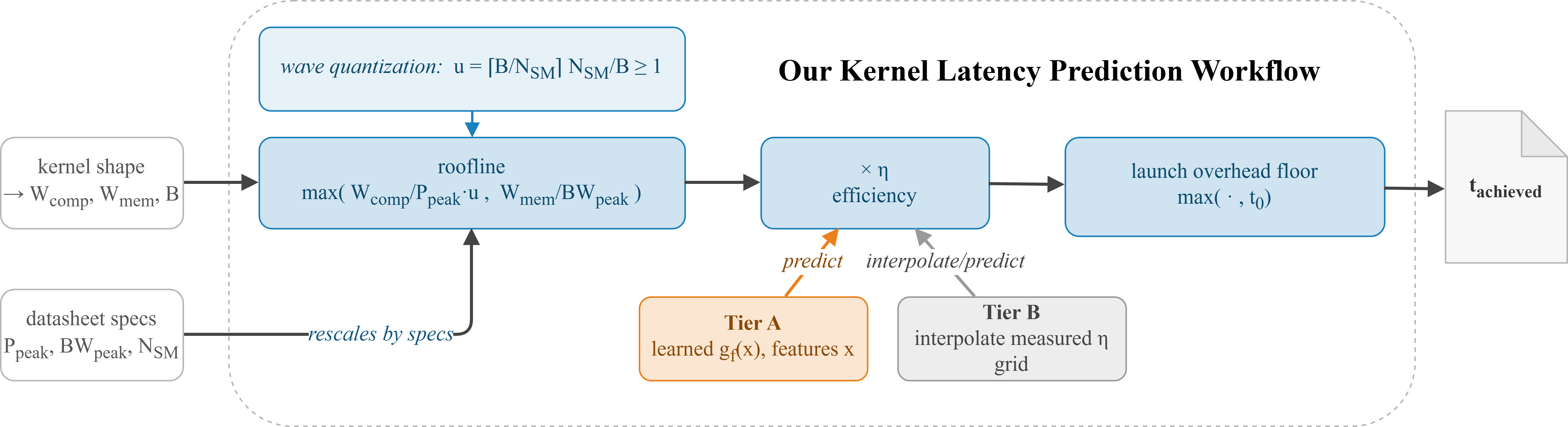}
  \caption{Our per-kernel latency prediction workflow.}
  \label{fig:pipeline}
\end{figure*}

\section{Kernel Latency Prediction Workflow}

\subsection{Tiered Prediction Architecture}
\label{sec:tiers}

A predictor is consulted at different points in a GPU's lifecycle with different information available: before the device exists, only its datasheet is known; a device in hand can be characterized once by a microbenchmark sweep, though the production model and workload may still be unknown. We therefore organize prediction into two \emph{tiers}, each a declared data policy: \textbf{Tier~A} uses the datasheet only (zero-shot cross-generation, \S\ref{sec:tierA}), and \textbf{Tier~B} adds a target microbenchmark grid and on-device calibration (\S\ref{sec:tierB}). The two are one system, not two: both use the same roofline$\times$efficiency formula (Eq.~\eqref{eq:master}) and differ only in where each factor comes from, so more target data simply sharpens the same predictor.

\subsection{Production Kernel Microbenchmarking}

Microbenchmarking, profiling individual kernels in isolation~\cite{jia_dissecting_2018, sun_dissecting_2023}, enables precise characterization of each operation's efficiency across the parameter space. Unlike NeuSight, which profiles generic operations, KernelSight-LM profiles the actual fused attention kernels in FlashAttention~\cite{dao_flashattention_2022a, dao_flashattention2_2023, shah_flashattention3_2024}. Since FlashAttention is widely deployed by production serving systems, including vLLM~\cite{kwon_efficient_2023a}, TGI~\cite{huggingface_text_2023}, and TensorRT-LLM~\cite{nvidiacorporation_tensorrtllm_2023}, and supported by the community, it contains many optimizations like fused kernels. This is critical because fused attention exhibits execution patterns and optimizations~\cite{shazeer_fast_2019, ainslie_gqa_2023} invisible to individual kernel profiling. Profiling within a production framework lets us capture these higher-level optimizations that isolated kernel profiling misses.

Furthermore, KernelSight-LM explicitly models the distinct serving phases: compute-bound prefill versus memory-bound decode, with asymmetric batch scaling. These phase-specific characteristics are fundamental to LLM inference and require separate predictors with phase-aware roofline formulations—a distinction that unified BMM modeling cannot capture.

\subsection{Efficiency Prediction}
\label{sec:eff-pred}

We model each kernel's latency as a hardware lower bound, scaled by a learned efficiency factor and floored by a fixed launch cost:

\begin{equation}
t = \max\!\big(t_{\text{roof}}\cdot\eta,\; t_0\big),
\qquad
t_{\text{roof}} = \max\!\Big(\tfrac{F_{\text{comp}}}{P_{\text{peak}}}\,u,\;
\tfrac{F_{\text{mem}}}{BW_{\text{peak}}}\Big),
\label{eq:master}
\end{equation}

where $F_{\text{comp}}$ and $F_{\text{mem}}$ are the kernel's FLOPs and bytes,
$P_{\text{peak}}$ and $BW_{\text{peak}}$ are the device peak compute and bandwidth,
$u=\lceil B/N_{\mathrm{SM}}\rceil\,N_{\mathrm{SM}}/B \ge 1$ is a wave-quantization
correction (the partial final wave when the kernel's $B$ blocks do not evenly fill
the $N_{\mathrm{SM}}$ SMs)~\cite{_matrix_}, and $t_0$ is the launch floor. The efficiency factor
$\eta = t_{\text{meas}}/t_{\text{roof}} \ge 1$ is the ratio of achieved to roofline
latency. $\eta\!=\!1$ denotes a kernel that attains its bound, and larger $\eta$ a
proportionally larger gap. Because $t_{\text{roof}}$ is recomputed entirely from
device specifications, it absorbs all hardware scaling, leaving $\eta$ as the only
quantity that must be measured or transferred (Figure~\ref{fig:pipeline}). The launch
floor is applied \emph{after} $\eta$, as a hard lower bound: a control-plane launch
cost is not an execution-throughput phenomenon and must not be scaled by an
efficiency factor.

\paragraph{Per-family heads on dimensionless features.}
We learn $\eta$ separately for each kernel family $f$. This includes GEMM, the LM head, attention
prefill and decode, and the elementwise and memory kernels. The formula is $\hat{\eta} =
g_f(\mathbf{x})$. The features $\mathbf{x}$ are deliberately \emph{dimensionless}:
arithmetic intensity and its position relative to the roofline ridge, per-SM work
and occupancy, wave count, and L2 cache residency. Being ratios, they describe a
kernel's operating point on the roofline independently of the device, which is what
allows a model fit on one set of GPUs to apply to another. Each head is trained against
$\log\eta=\log(t_{\text{meas}}/t_{\text{roof}})$, where $t_{\text{meas}}$ is the
CUDA-graph microbenchmark body.

\begin{figure}[t]
  \centering
  \includegraphics[width=\columnwidth]{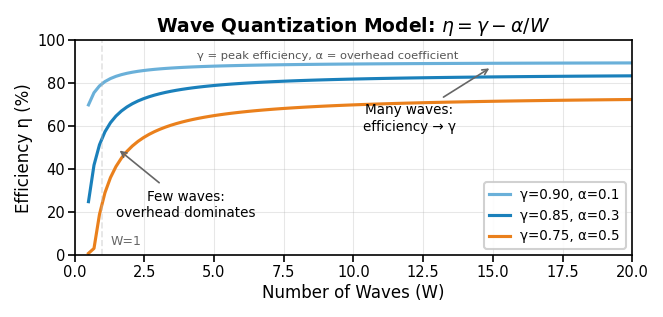}
  \caption{The bounded analytical head adapted from NeuSight, plotted as achieved
  efficiency versus wave count $W$ for three $(\gamma,\alpha)$ settings: efficiency
  climbs toward the peak $\gamma$ as the partial-wave overhead $\alpha/W$ decays. We
  retain it as one admissible per-family head; in Eq.~\eqref{eq:master} the efficiency
  \emph{factor} is the reciprocal of this utilization,
  $\eta=(\gamma-\alpha/W)^{-1}\!\ge\!1$.}
  \label{fig:efficiency_factor}
\end{figure}

\paragraph{A diverse, low-capacity head family.}
Cross-generation prediction is an extrapolation from a small number of training
devices, a regime in which an appropriate inductive bias outperforms model capacity.
We therefore draw each per-family head from a diverse, low-capacity pool, a
regularized linear model, gradient-boosted and bagged trees, a small MLP ensemble,
and the bounded analytical form $\eta = \big(\gamma(\mathbf{x}) -
\alpha(\mathbf{x})/W\big)^{-1}$ adapted from NeuSight~\cite{lee_forecasting_2025},
where $W$ is the wave count (Figure~\ref{fig:efficiency_factor}). We then select per family
by cross-validation over held-out source devices. The bounded form imposes a monotone
wave-quantization prior and is selected where that prior regularizes better than a
free model (compute-bound kernels and smooth memory-bound writes). Flexible heads are
selected where the efficiency surface is non-monotone, such as attention. Our
framework thus \emph{contains} the bounded form as one admissible head rather than
fixing it for every operator. Wave quantization is modeled in the roofline (the $u$
term of Eq.~\ref{eq:master}) rather than folded into $\eta$, which keeps the learned
residual small and smooth. The extent to which $\eta$ is genuinely transferable
varies by family and is examined in \S\ref{sec:limitations}.

\subsection{Tier A: Zero-Shot Cross-Generation Prediction}
\label{sec:tierA}

Tier~A predicts a GPU that has never been benchmarked: only its datasheet is
available. The datasheet should provide us with peak compute $P_{\text{peak}}$, peak bandwidth $BW_{\text{peak}}$, SM count, and cache sizes. This is the setting for capacity planning and for forecasting an unreleased or unavailable generation. The roofline is evaluated directly from the target's datasheet peaks, so it
rescales to the new device by substitution. The efficiency factor
is supplied by the learned model $g_f(\mathbf{x})$ of \S\ref{sec:eff-pred}, trained on
a ladder of \emph{source} GPUs with the target device excluded. The launch floor
$t_0$ is treated as a portable constant. As no host-side calibration is available
without target traces, the host terms of the step assembly (\S\ref{sec:assembly})
are set to the calibration value obtained from its closest relative. Tier~A therefore
predicts the device-scaled step body.

\subsection{Tier B: Target-Measured Prediction}
\label{sec:tierB}

Tier~B applies when the target GPU is available and can be characterized once. A
single characterization pass yields two products. First, a kernel microbenchmark
sweep gives \emph{measured} peaks for the roofline and a dense grid of measured
efficiency values; for a query shape, $\eta$ is obtained by interpolating this grid
over the dimensionless feature coordinates rather than from the learned model. We
interpolate the bounded residual $\eta$ rather than raw latency, because the roofline
already accounts for the order-of-magnitude scaling across shapes, leaving a smooth,
well-conditioned quantity to interpolate. Where a query lies outside the measured
convex hull, the predictor falls back to the learned model of \S\ref{sec:eff-pred}.
The predictor is thus \emph{density-routed}. It uses measured interpolation in the dense
interior, learned extrapolation at the edge. This makes it more accurate than Tier~A.

Second, the same characterization runs a short serving sweep to measure the host-side
constants of the step assembly (\S\ref{sec:assembly}): the CPU overlap and serial-tail
floors, request admission, eager-prefill launch, and the microbenchmark-to-serving
body scale. These cannot be obtained from the kernel microbenchmark, which does not
exercise the scheduler, sampler, or request lifecycle. Being properties of the
$(\text{host CPU},\text{GPU},\text{engine})$ stack rather than the GPU alone, they are
re-measured whenever the host or serving engine changes; the measured peaks and the
efficiency grid, in contrast, transfer with the GPU.

\subsection{Communication Latency}
\label{sec:comm}

Tensor-parallel inference inserts an all-reduce after the attention and MLP blocks
of each layer to combine partial activations across the $P$ shards~\cite{shoeybi_megatronlm_2020a}. Communication
follows the same bandwidth-bound structure as the kernel roofline, but on the interconnect rather than HBM. We model the ring all-reduce~\cite{patarasuk_bandwidth_2009} as

\begin{equation}
    t_{\text{ar}} = \alpha_0 + \frac{2(P-1)}{P}\cdot\frac{S}{\beta_{\text{eff}}},
    \label{eq:allreduce}
\end{equation}

where $S$ is the message size in bytes, $\beta_{\text{eff}}$ the effective
interconnect bandwidth, and $\alpha_0$ a startup/synchronization term that grows weakly
with $P$. The factor $2(P-1)/P$ is the per-GPU volume of the reduce-scatter and
all-gather passes. Each layer contributes two such all-reduces to $t_{\text{layer}}$
(\S\ref{sec:assembly}).

\begin{figure}
  \centering
  \includegraphics[width=0.95\columnwidth]{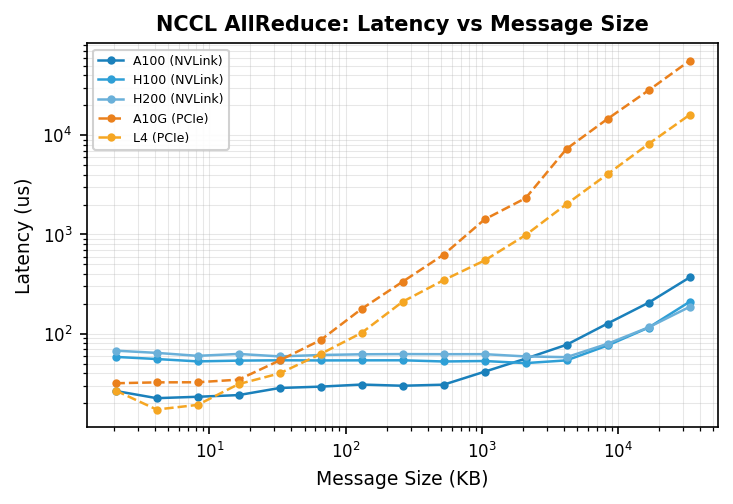}  
  \caption{NVLink and PCIe differ in absolute bandwidth, but their latency follows the same form; we fit one common predictor to both and predict latency from the formula.}
  \label{fig:comm_latency}
\end{figure}

Consistent with the prediction tiers, $\beta_{\text{eff}}$ is the datasheet NVLink
bandwidth for an unbenchmarked target (Tier~A) and a measured value when available
(Tier~B). Equation~\eqref{eq:allreduce} fits measured NCCL data~\cite{_nvidia_2026} closely
(Figure~\ref{fig:comm_latency}). We fit $\alpha_0$ and $\beta_{\text{eff}}$ per
interconnect where collective measurements exist, and fall back to the datasheet
bandwidth otherwise.

\begin{figure*}
  \centering
  \includegraphics[width=0.95\textwidth]{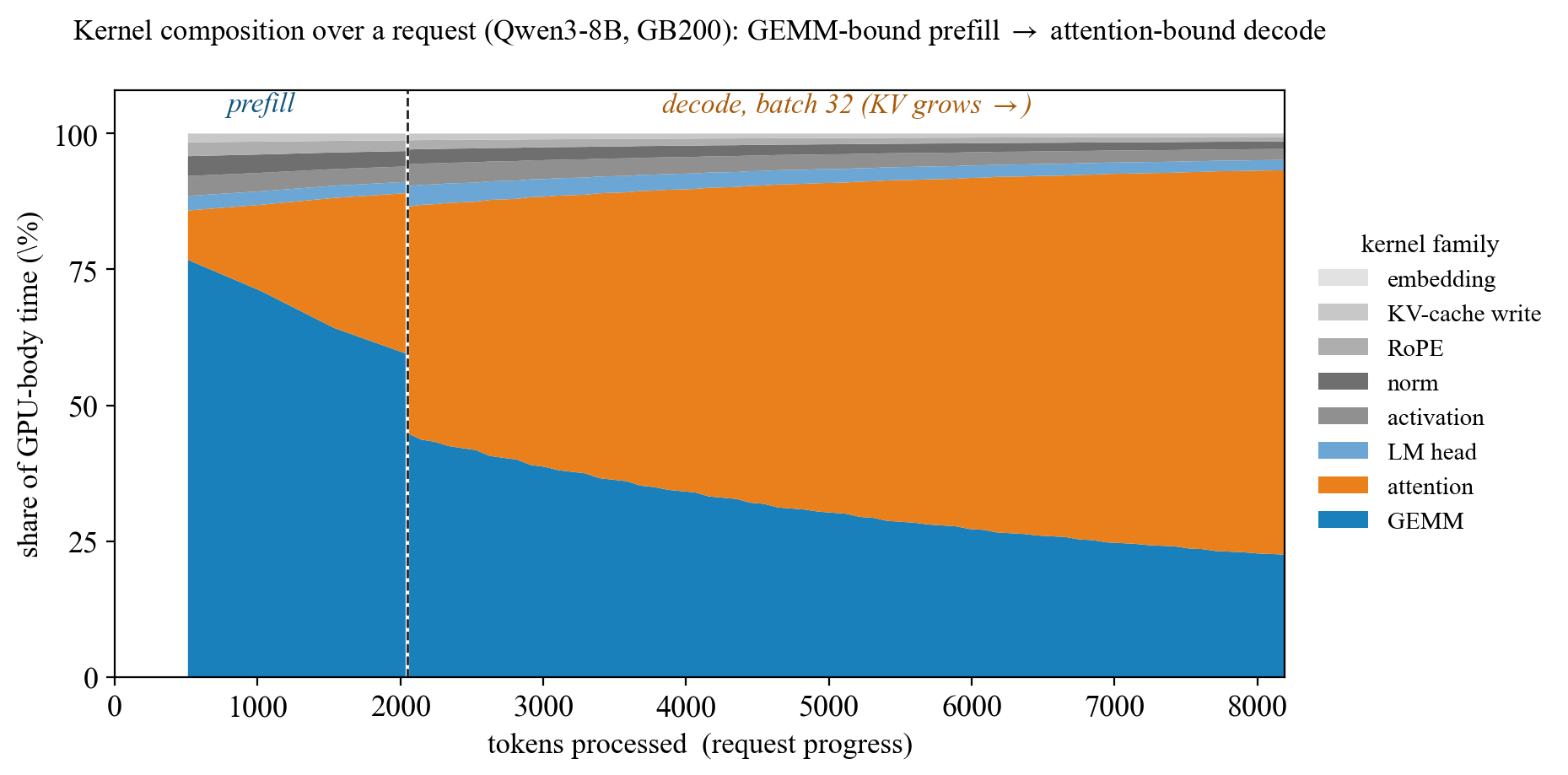}  
\caption{Kernel-time composition over an example single request's progress (Qwen3-8B, GB200). During prefill, GEMM dominates ($\sim$77\%), but when decode begins, the matmuls shrink to $M\!=\!\text{batch}$, and GEMM's share drops sharply, while attention grows with the KV length to dominate at long context.}
\label{fig:composition-progress}

\end{figure*}

\subsection{Execution Latency Aggregation and Overhead Calibration}
\label{sec:assembly}

Per-kernel predictions are aggregated along the transformer hierarchy: a layer's
operator kernels $t_{\text{kernel}}=\max(t_{\text{roof}}\,\eta,\,t_0)$ are summed with
its tensor-parallel all-reduce (\S\ref{sec:comm}); since decoder layers are
homogeneous, the GPU body of a step is $t_{\text{gpu}}=L\,t_{\text{layer}}+
t_{\text{lm\_head}}+t_{\text{embed}}$. The step wall time is not simply
$t_{\text{gpu}}$, because the serving engine overlaps host work (launch, sampling,
scheduling) with GPU compute:
\begin{equation}
t_{\text{step}} = \max\!\big(t_{\text{gpu}}+n\,c_{\text{disp}},\ \ell\big)
+ \max\!\big(0,\ c_{\text{ovl}}-t_{\text{gpu}}\big) + c_{\text{tail}},
\label{eq:assembly}
\end{equation}
where the host launch/overlap floor $c_{\text{ovl}}$ surfaces only when it exceeds the
GPU and the serial tail $c_{\text{tail}}$ is always on the critical path. an additive
model would over-count, since for small models the host pipeline, not the GPU, bounds
the step.

These overhead constants depend on the host CPU and serving engine, not on the GPU, so
we measure them once from a short serving sweep during Tier~B characterization
(\S\ref{sec:tierB}); Tier~A uses a simple formula to get an estimate based on its close relatives' calibration numbers. Substituting measured per-step times for the predictor in Eq.~\eqref{eq:assembly} reproduces real vLLM latency to within $\approx$2\%,
confirming residual error stems from per-kernel prediction, not aggregation.

\subsection{Extended Simulator Capabilities}

In our scheduler, we follow the same mechanism as Vidur~\cite{agrawal_vidur_2024}. At each step, the scheduler emits a batch descriptor (decode tokens, prefill-chunk lengths, per-request KV lengths) that is exactly the predictor's input, so scheduling fidelity is a prerequisite for accurate latency. Beyond this, we also made some changes to Vidur's simulator framework to make it support more advanced inference scenarios.

\textbf{Faithful vLLM~v1 scheduling.} We re-implement vLLM~v1's scheduler rather than
approximating it: a single token-budget loop advances every request's computed-token
count toward its target, which unifies continuous batching, chunked
prefill, and prefix-cache-aware admission in one algorithm, with FCFS queuing and
preemption/recompute under KV pressure. We validated it line-by-line against the
vLLM~v1 source, so the simulated batch composition tracks the real engine step-for-step. The request launch sequence and time in simulated and real traces achieve a Spearman correlation score of $+0.985$.

\textbf{Hierarchical prefix caching.} We model vLLM's content-hashed KV cache~\cite{zheng_sglang_2024, gim_prompt_2024} in two
tiers: a volatile GPU tier and a persistent disk tier~\cite{kwon_efficient_2023a, gao_costefficient_2024, qin_mooncake_2025}. A
prefix hit reuses cached blocks so only the uncached suffix is recomputed, lowering
predicted latency through both less prefill work and more concurrent requests from
freed KV capacity.

\textbf{Mixed prefill-decode steps.} Continuous batching co-schedules decoding
requests with a prefilling request's token chunk in a single forward pass~\cite{agrawal_sarathi_2023, agrawal_taming_2024}. We price
such mixed steps kernel-by-kernel from the step's token composition: the dense
GEMMs, norms, and activations run over the concatenated tokens (priced at the total
token count), while attention is computed per request, each decode over its cached
context, and the prefill chunk over its cached history plus its own causal
triangle. The host launch term adds a measured per-layer cost, with a
correction fitted based on traces to add penalties for smaller chunks. Because the prediction is assembled from the actual per-step token counts rather than looked up, it covers the combinatorial space of mixed-step shapes that a profiled operator table cannot enumerate.

\section{Experimental Setup}

In our evaluation, we profile graphed GPU kernels on a select number of GPUs and evaluate cross-generation prediction on an unseen target. We compare simulated serving metrics against vLLM profiled ground truth on a production conversation trace, up to 500 requests per configuration, across six model families with models from 0.5B to 70B parameters.

\paragraph{Baselines and scenarios.} For Tier~A, we compare our per-kernel accuracy with NeuSight~\cite{lee_forecasting_2025}.

For Tier~B and end-to-end inference, we compare against the profiling-based Random Forest (RF) predictors of Vidur~\cite{agrawal_vidur_2024} and AIConfigurator's prediction workflow.

We also tested Vidur's Linear Regression (LR) model on the data, but its error rate is too high to be considered relevant.

\paragraph{GPU selection.}
We profile every device in Table~\ref{tab:gpu_specs} but train and evaluate on only a
curated subset. We omit the oldest and lowest-end SKUs: their efficiency is governed by
regimes (small caches, low occupancy, narrow memory buses) far from modern targets, so
using them as source devices adds noise to the learned efficiency factor without
improving transfer to recent hardware. The cross-generation ladder is therefore a
compact, datacenter-class set spanning Ampere, Hopper, and Ada (A100, H200-NVL, L40S),
with the Blackwell GB200 held out as the target. We also profile two edge accelerators,
the consumer RTX~5090 (GDDR7) and the GB10/RTX~Spark (LPDDR5X over NVLink-C2C): their
memory technology and compute-to-bandwidth balance differ sharply from datacenter
parts, so they are not directly comparable and are kept out of the ladder and the
headline target. They remain useful for analysis.

\section{Results and Analysis}

\subsection{Cross-Generation Kernel Prediction}
\label{sec:crossgen-results}

\begin{table}[t]
\centering
\caption{Cross-generation per-kernel accuracy: predicting NVIDIA GB200 based on prior-gen profiling data (MAPE, lower is better). Both methods use the same datasheet-driven roofline and differ only in the efficiency model. Training dataset: A100, H200-NVL, L40S.}
\label{tab:crossgen}
\begin{tabular}{l r r r}
\toprule
\textbf{Kernel family} & \textbf{NeuSight} & \textbf{Ours} & \textbf{Improv.} \\
\midrule
\multicolumn{4}{l}{\textit{Compute-bound (both near the cross-gen floor)}} \\
GEMM                & 13.4 & 12.8 & 1.0$\times$ \\
LM head             & \phantom{0}7.8 & \phantom{0}7.7 & 1.0$\times$ \\
\midrule
\multicolumn{4}{l}{\textit{Memory-bound \& attention (fixed form cannot adapt)}} \\
RMSNorm             & 15.9 & 13.1 & 1.2$\times$ \\
SiLU$\times$Mul     & 21.1 & 13.6 & 1.6$\times$ \\
RoPE                & 36.7 & 18.0 & 2.0$\times$ \\
KV-cache write      & 18.0 & \phantom{0}8.2 & 2.2$\times$ \\
Attention (prefill) & 37.1 & 13.6 & 2.7$\times$ \\
Attention (decode)  & 26.3 & \phantom{0}9.3 & 2.8$\times$ \\
\midrule
\textbf{Mean}       & 22.0 & 12.1 & 1.8$\times$ \\
\bottomrule
\end{tabular}

\end{table}

\begin{table*}\centering\small\setlength{\tabcolsep}{4pt}
\caption{GB200 \emph{per-request} error-distribution: median (p50), bad (p90), worst-percentile (p99)
and worst-single-request (max) of $|$APE$|$, pooled over all requests in each cut, as Tier~A\,/\,Tier~B\,/\,AIConfigurator. Worst cases are long-context Qwen3-8B requests
(input sequence length $\approx$19--23k). Vidur is omitted from the table (see text).}

\label{tab:percut_errdist}
\begin{tabular}{l l cccc cccc}
\toprule
& & \multicolumn{4}{c}{\textbf{TTFT} (Tier~A/Tier~B/AIC)} & \multicolumn{4}{c}{\textbf{TPOT} (Tier~A/Tier~B/AIC)} \\
\cmidrule(lr){3-6}\cmidrule(lr){7-10}
\textbf{Cut} & \textbf{Group} & p50 & p90 & p99 & max & p50 & p90 & p99 & max \\
\midrule
\multirow{3}{*}{Family}
 & Qwen3    & 19/17/55 & 52/52/78 & 80/77/83 & 149/141/91 & 13/5/7  & 29/17/11 & 39/26/26 & 92/33/35 \\
 & Llama    & 13/13/74 & 27/26/86 & 81/81/89 & 90/90/92   & 16/7/4  & 32/16/8  & 34/27/13 & 36/33/17 \\
 & DeepSeek & 13/13/59 & 32/32/80 & 88/88/85 & 96/96/85   & 9/7/4   & 15/12/7  & 28/13/8  & 30/13/10 \\
\midrule
\multirow{3}{*}{Size}
 & small & 19/18/74 & 51/51/85 & 89/89/90 & 149/141/92 & 20/6/5 & 32/20/9 & 38/29/26 & 92/33/35 \\
 & mid   & 13/12/55 & 33/32/73 & 70/70/78 & 79/77/83   & 8/6/7  & 18/9/11 & 23/12/14 & 26/15/17 \\
 & large & 11/12/47 & 24/25/64 & 46/46/66 & 66/67/66   & 6/8/5  & 10/12/7 & 13/13/8  & 16/13/9 \\
\midrule
\multirow{3}{*}{TP}
 & tp1 & 22/22/63 & 57/56/87 & 94/94/89 & 149/141/89 & 26/5/4  & 33/9/9  & 40/21/30 & 92/26/35 \\
 & tp2 & 13/13/56 & 33/31/80 & 79/78/85 & 81/81/90   & 8/3/5   & 13/8/8  & 20/10/14 & 23/11/17 \\
 & tp4 & 14/12/62 & 29/27/82 & 69/69/87 & 81/81/92   & 13/10/7 & 27/21/10& 33/30/16 & 36/33/17 \\
\midrule
\textbf{All} & & 15/14/59 & 41/41/83 & 81/81/89 & 149/141/92 & 12/6/5 & 29/14/9 & 36/27/17 & 92/33/35 \\

\bottomrule
\end{tabular}
\end{table*}

We evaluate Tier~A by predicting GB200 with the device fully held out from
training (Table~\ref{tab:crossgen}), against NeuSight~\cite{lee_forecasting_2025}. Both methods use the identical datasheet roofline and the same held-out ladder, so the table isolates the one thing they differ on: the efficiency model.

\paragraph{Compute-bound kernels}
GEMM and the LM head are a tie (within $\sim$2\,pp). Their efficiency is smooth
and only weakly device-dependent, so NeuSight's single bounded form already
captures it, and both methods sit near the irreducible cross-generation floor.
Extra capacity does not help much here.

\paragraph{Memory-bound and attention kernels}
NeuSight applies a single bounded efficiency shape to every kernel. The
memory-bound families (RMSNorm, RoPE, activation, KV-cache) and attention have
efficiencies that vary strongly with shape, driven by achieved-bandwidth ceilings, occupancy, and cache residency rather than a single wave term, which one fixed form cannot track. Our per-family heads adapt to each, cutting error $1.2$--$2.8\times$. This block accounts for the majority of the mean improvement.

\paragraph{Attention and kernel identity.}
Decode attention additionally requires matching the kernel the engine dispatches:
Predicting it from the served kernel rather than an isolated benchmark cuts error
from $26.3$ to $9.3$ ($2.8\times$ over NeuSight). This is the same kernel-identity
property formalized in \S\ref{sec:tierA}. Without it, an operator-keyed
baseline trains on a kernel the target never runs.

Another issue is that some portion of the efficiency remains device-specific, and thus untransferable. We discuss this in \S\ref{sec:limitations}.

\paragraph{Changing Composition of Kernel runtime}

As shown in Table~\ref{tab:crossgen}, our predictor's advantage concentrates in the memory-bound and attention kernels. In real traces, however, GEMM accounts for a large share of kernel runtime, so the realized end-to-end improvement may be smaller than the per-kernel mean suggests.

As a request progresses, the kernel composition shifts substantially. Figure~\ref{fig:composition-progress} shows an example. This is why simple calculation based on a fixed kernel decomposition is inaccurate.

\subsection{Accurate In-Distribution Kernel Latency Prediction}
\label{sec:why-kernel-wins}

\begin{table}[t]
\centering
\caption{In-distribution per-kernel accuracy on NVIDIA GB200 (MAPE, lower is better). \textsc{KernelSight-LM} uses Tier~B (\S\ref{sec:tierB}): per-shape efficiency interpolated over the measured grid. \textsc{AIConfigurator} either interpolates a measured \emph{operator}-level table or, for the elementwise/memory kernels, falls back to a fixed analytical model (\texttt{query\_mem\_op}~$=\text{bytes}/(0.8\,\text{BW})+3\,\mu s$).}

\label{tab:kernel-comparison}
\begin{tabular}{l r r r}
\toprule
\textbf{Kernel family} & \textbf{AIConfig.} & \textbf{Ours} & \textbf{Improv.} \\
\midrule
\multicolumn{4}{l}{\textit{Operators they measure (table interpolation)}} \\
GEMM                   & \phantom{0}2.8 & 2.6 & $1.1\times$ \\
Attention (decode)     & 11.3 & 3.6 & $3.1\times$ \\
Attention (prefill)    & 21.3 & 9.4 & $2.3\times$ \\
LM head                & 16.2 & 4.7 & $3.4\times$ \\
\midrule
\multicolumn{4}{l}{\textit{Kernels they model analytically (\texttt{query\_mem\_op})}} \\
RMSNorm                & 29.9 & 2.7 & $11\times$ \\
RoPE                   & 31.7 & 3.8 & $8.3\times$ \\
SiLU$\times$Mul (act.) & 29.0 & 3.5 & $8.3\times$ \\
KV-cache write         & 43.1 & 3.3 & $13\times$ \\
Embedding              & 64.1 & 0.9 & $71\times$ \\
\midrule
\textbf{Mean}          & 27.7 & 3.8 & $7.3\times$ \\
\bottomrule
\end{tabular}
\end{table}

Table~\ref{tab:kernel-comparison} compares Tier~B against \textsc{AIConfigurator}~\cite{xu_aiconfigurator_2026} on GB200. The advantage comes from two
distinct mechanisms, corresponding to the two ways an operator-level system
loses resolution.

\paragraph{Completeness: measured kernels vs.\ an analytical shortcut.}
Because \textsc{AIConfigurator} optimizes deployment \emph{configurations}, where
GEMMs and attention dominate runtime, it does not measure the small elementwise
and memory kernels at all. RMSNorm, RoPE, the activation, the KV-cache write, and
the embedding are routed through a single closed form, $\text{bytes}/(0.8\,\text{BW})+3\,\mu s$. This bakes in one fixed achieved-bandwidth fraction ($0.8$) and one fixed launch floor ($3\,\mu s$) for every kernel and shape; the measured ground truth violates both (per-family bias $0.73$--$1.64\times$). By measuring these kernels and learning a per-shape efficiency, \textsc{KernelSight-LM} cuts their error from $29$--$64\%$ to under $4\%$, a $\sim$$10\times$ gain that exists purely because we collect data the operator-level system never does.

\paragraph{Granularity: the efficiency target.}
For the operators \textsc{AIConfigurator} \emph{does} measure, the remaining
advantage comes from \emph{what} is interpolated. As described in
\S\ref{sec:tierB}, we let the roofline carry the order-of-magnitude scale
and interpolate only the bounded efficiency residual, which is far smoother than
a raw-latency surface that ranges over the LM head's token count or attention's
sequence and KV length. Both systems use the same measured data, yet attention
and the LM head still improve $2$--$3.5\times$. GEMM is the lone tie: a single,
well-conditioned kernel whose latency surface is already smooth, so raw-latency
interpolation is as good as ours.

\paragraph{Accuracy as a symptom.}
These gains are a symptom of the deeper design difference established in
\S\ref{sec:tierB}: kernel-level prediction records the kernel that
actually ran and is strictly more expressive. It aggregates up to any operator,
recomposes unmeasured configurations, and (\S\ref{sec:tierA}) transfers across
GPU generations, none of which an operator table supports.

\paragraph{Translation to end-to-end improvement.}
Per-kernel accuracy does not translate one-to-one end-to-end: AIConfigurator's on-device, multi-trace calibration closes much of the decode-body gap there (\S\ref{sec:end-to-end}), where the two are comparable.

\begin{figure*}[t]
  \centering
  \includegraphics[width=\textwidth]{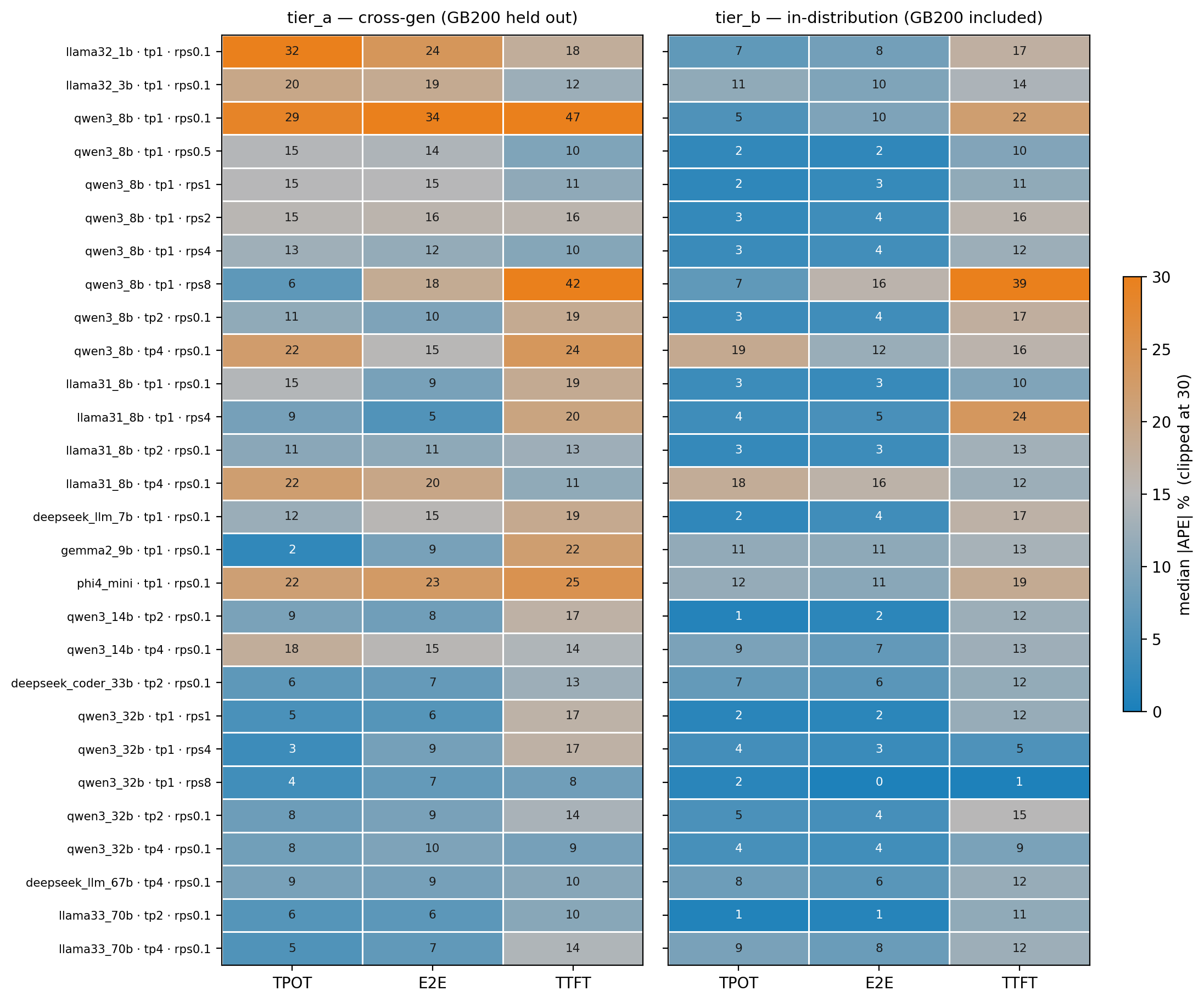}
  \caption{GB200 real-vs-sim accuracy across select serving runs
    (models $\times$ TP $\times$ request rate) and three metrics. Tier~A is zero-shot (cross-gen efficiency model plus a
    portable launch overhead, no GB200 host data); Tier~B adds
    target-device microbenchmarks and a clean per-model host calibration.
    Each row is one $(\text{model},\text{TP},\text{request-rate})$ run and each column one serving metric; cell color encodes the median absolute percent error for that run and metric, clipped at 30\% (darker = larger error).}
  \label{fig:calibrated-accuracy}
\end{figure*}

\subsection{End-to-end Inference Performance Prediction}

\label{sec:end-to-end}

Table~\ref{tab:percut_errdist} reports the \emph{distribution} of per-request prediction
error on GB200 across various serving runs spanning different model families, sizes, and tensor
parallelism, and Figure~\ref{fig:calibrated-accuracy} shows the corresponding per-run accuracy heatmap. For each cut, we give the median request error (p50) and the tail (p90 and p99); this shows more clearly how the prediction mechanisms compare. We evaluate two tiers of KernelSight-LM against AIConfigurator; Vidur is omitted and discussed separately below. KernelSight-LM is the most accurate while using the least target data in almost all cases.

\textbf{TPOT} Decode latency is predicted tightly. Tier~B's typical request is within 6\%
and its worst within 33\%; even the zero-shot Tier~A holds a 12\% median (36\% at p99).
AIConfigurator is slightly better than ours in the decode body (5\% median), likely because AIConfigurator is calibrated on-device from multiple run traces, while our Tier~B uses only a short microbenchmark.

Vidur collapses to a 436\% median error: its profiles use a different mode than actual vLLM runs, so it predicts a flat decode step several times the real graphed latency in every case. We omit Vidur from the table, as its predictions are uniformly far from the measured values.

\textbf{TTFT} Time-to-first-token is harder. Our predictors track the typical request well
(14--15\% median) but have a long tail (81\% at p99). The worst cases are all long-context
prefills (19--23k tokens), where attention is the least-transferable kernel. The gap to the
external estimators is sharpest here, and it is \emph{structural}, not a tail effect:
AIConfigurator's 59\% median and Vidur's 56\% median both stem from having no host-overhead model, which makes their TTFT predictions effectively unusable.

Short prompts are dominated by admission, framing, and launch overhead, while long, compute-bound prompts are dominated more by GPU compute. KernelSight-LM represents these overheads explicitly in its discrete-event composition, making its typical-case TTFT roughly $4\times$ more accurate than either baseline.

\textbf{Throughput} Output throughput is the metric KernelSight-LM predicts most
accurately (p50 at 3.0\% MAPE). More specifically, the error is below 1\% under light load, where throughput is arrival-limited, and rises to only 8\% at saturation, where the simulator must reproduce the server's true sustained rate. 

AIConfigurator can achieve similar accuracy (2.7\% at p50) at light load or below saturation. Saturation is exactly where AIConfigurator's design breaks down: as a GPU-bound capacity estimator, it lacks the mechanism to model host and scheduling in detail, so the error rate explodes.

\section{Conclusions}

We presented KernelSight-LM, a kernel-level latency prediction framework for LLM serving simulation that achieves cross-GPU generalization without requiring per-device profiling. By decomposing transformer execution into fundamental kernels, our approach predicts latencies on unseen GPUs using only their published specifications or, more accurately, with a small on-device microbenchmarking. Integrated with an execution simulator, KernelSight-LM enables rapid evaluation of serving configurations across diverse hardware by providing high-fidelity predictions on multiple metrics. Our experiments demonstrate that kernel-level prediction achieves better accuracy than profiling-based methods. Our simulator extensions further enable exploring a wider search space of optimal configurations and supporting hardware-software co-design.

Looking ahead, the main open problem is the limited cross-generation transfer of the attention efficiency factor, which motivates one-shot per-architecture attention calibration; further directions include extending the host model into the saturated regime and broadening interconnect coverage.

\FloatBarrier

\bibliographystyle{ACM-Reference-Format}

\bibliography{citations, software}

@misc{_matrix_,
  title = {Matrix {{Multiplication Background User}}'s {{Guide}}},
  author = {{NVIDIA Corporation}},
  year = {2023},
  howpublished = {NVIDIA Deep Learning Performance Documentation, \url{https://docs.nvidia.com/deeplearning/performance/dl-performance-matrix-multiplication/index.html}},
  note = {Accessed: 2026-06-25}
}

@misc{_nvidia_2026,
  title = {{{NCCL}}: Optimized Primitives for Collective Multi-{{GPU}} Communication},
  author = {{NVIDIA Corporation}},
  year = {2024},
  howpublished = {\url{https://github.com/NVIDIA/nccl}},
  note = {Accessed: 2026-06-25}
}

@inproceedings{_orca_,
  title = {Orca: {{A Distributed Serving System}} for {{Transformer-Based Generative Models}}},
  author = {Yu, Gyeong-In and Jeong, Joo Seong and Kim, Geon-Woo and Kim, Soojeong and Chun, Byung-Gon},
  booktitle = {16th {{USENIX}} Symposium on Operating Systems Design and Implementation ({{OSDI}} 22)},
  year = {2022},
  pages = {521--538},
  isbn = {978-1-939133-28-1},
  url = {https://www.usenix.org/conference/osdi22/presentation/yu}
}

@misc{abdin_phi4_2024,
  title = {Phi-4 {{Technical Report}}},
  year = 2024,
  month = dec,
  number = {arXiv:2412.08905},
  eprint = {2412.08905},
  primaryclass = {cs.CL},
  publisher = {arXiv},
  doi = {10.48550/arXiv.2412.08905},
  urldate = {2026-06-25},
  archiveprefix = {arXiv},
  author = {Abdin, Marah and Aneja, Jyoti and Behl, Harkirat and Bubeck, S{\'e}bastien and Eldan, Ronen and Gunasekar, Suriya and Harrison, Michael and Hewett, Russell J. and others}
}

@misc{agrawal_sarathi_2023,
  title = {{{SARATHI}}: {{Efficient LLM Inference}} by {{Piggybacking Decodes}} with {{Chunked Prefills}}},
  shorttitle = {{{SARATHI}}},
  author = {Agrawal, Amey and Panwar, Ashish and Mohan, Jayashree and Kwatra, Nipun and Gulavani, Bhargav S. and Ramjee, Ramachandran},
  year = 2023,
  month = aug,
  number = {arXiv:2308.16369},
  eprint = {2308.16369},
  primaryclass = {cs.LG},
  publisher = {arXiv},
  doi = {10.48550/arXiv.2308.16369},
  urldate = {2026-06-25},
  archiveprefix = {arXiv}
}

@misc{agrawal_taming_2024,
  title = {Taming {{Throughput-Latency Tradeoff}} in {{LLM Inference}} with {{Sarathi-Serve}}},
  author = {Agrawal, Amey and Kedia, Nitin and Panwar, Ashish and Mohan, Jayashree and Kwatra, Nipun and Gulavani, Bhargav S. and Tumanov, Alexey and Ramjee, Ramachandran},
  year = 2024,
  month = jun,
  number = {arXiv:2403.02310},
  eprint = {2403.02310},
  primaryclass = {cs.LG},
  publisher = {arXiv},
  doi = {10.48550/arXiv.2403.02310},
  urldate = {2026-06-25},
  archiveprefix = {arXiv}
}

@article{agrawal_vidur_2024,
  title = {Vidur: {{A}} Large-Scale Simulation Framework for Llm Inference},
  author = {Agrawal, Amey and Kedia, Nitin and Mohan, Jayashree and Panwar, Ashish and Kwatra, Nipun and Gulavani, Bhargav S and Ramjee, Ramachandran and Tumanov, Alexey},
  year = 2024,
  journal = {Proceedings of Machine Learning and Systems},
  volume = {6},
  pages = {351--366}
}

@article{ainslie_gqa_2023,
  title = {Gqa: {{Training}} Generalized Multi-Query Transformer Models from Multi-Head Checkpoints},
  author = {Ainslie, Joshua and {Lee-Thorp}, James and De Jong, Michiel and Zemlyanskiy, Yury and Lebr{\'o}n, Federico and Sanghai, Sumit},
  year = 2023,
  journal = {arXiv preprint arXiv:2305.13245},
  eprint = {2305.13245},
  archiveprefix = {arXiv}
}

@article{bi_deepseek_2024,
  title = {Deepseek Llm: {{Scaling}} Open-Source Language Models with Longtermism},
  year = 2024,
  journal = {arXiv preprint arXiv:2401.02954},
  eprint = {2401.02954},
  archiveprefix = {arXiv},
  author = {Bi, Xiao and Chen, Deli and Chen, Guanting and Chen, Shanhuang and Dai, Damai and Deng, Chengqi and Ding, Honghui and Dong, Kai and others}
}

@inproceedings{brown_language_2020,
  title = {Language Models Are Few-Shot Learners},
  booktitle = {Proceedings of the 34th {{International Conference}} on {{Neural Information Processing Systems}}},
  year = 2020,
  month = dec,
  series = {{{NIPS}} '20},
  pages = {1877--1901},
  publisher = {Curran Associates Inc.},
  address = {Red Hook, NY, USA},
  urldate = {2026-01-28},
  isbn = {978-1-7138-2954-6},
  author = {Brown, Tom B. and Mann, Benjamin and Ryder, Nick and Subbiah, Melanie and Kaplan, Jared and Dhariwal, Prafulla and Neelakantan, Arvind and Shyam, Pranav and others}
}

@article{cho_llmservingsim20_2025,
  title = {{{LLMServingSim2}}.0: {{A Unified Simulator}} for {{Heterogeneous Hardware}} and {{Serving Techniques}} in {{LLM Infrastructure}}},
  shorttitle = {{{LLMServingSim2}}.0},
  author = {Cho, Jaehong and Choi, Hyunmin and Park, Jongse},
  year = 2025,
  month = jul,
  journal = {IEEE Computer Architecture Letters},
  volume = {24},
  number = {2},
  eprint = {2511.07229},
  primaryclass = {cs},
  pages = {361--364},
  issn = {1556-6056, 1556-6064, 2473-2575},
  doi = {10.1109/LCA.2025.3628325},
  urldate = {2026-01-29},
  archiveprefix = {arXiv}
}

@misc{dao_flashattention_2022a,
  title = {{{FlashAttention}}: {{Fast}} and {{Memory-Efficient Exact Attention}} with {{IO-Awareness}}},
  shorttitle = {{{FlashAttention}}},
  author = {Dao, Tri and Fu, Daniel Y. and Ermon, Stefano and Rudra, Atri and R{\'e}, Christopher},
  year = 2022,
  month = jun,
  number = {arXiv:2205.14135},
  eprint = {2205.14135},
  primaryclass = {cs.LG},
  publisher = {arXiv},
  doi = {10.48550/arXiv.2205.14135},
  urldate = {2026-06-25},
  archiveprefix = {arXiv}
}

@article{dao_flashattention2_2023,
  title = {Flashattention-2: {{Faster}} Attention with Better Parallelism and Work Partitioning},
  author = {Dao, Tri},
  year = 2023,
  journal = {arXiv preprint arXiv:2307.08691},
  eprint = {2307.08691},
  archiveprefix = {arXiv}
}

@misc{gao_costefficient_2024,
  title = {Cost-{{Efficient Large Language Model Serving}} for {{Multi-turn Conversations}} with {{CachedAttention}}},
  year = 2024,
  month = jun,
  number = {arXiv:2403.19708},
  eprint = {2403.19708},
  primaryclass = {cs.CL},
  publisher = {arXiv},
  doi = {10.48550/arXiv.2403.19708},
  urldate = {2026-06-25},
  archiveprefix = {arXiv},
  author = {Gao, Bin and He, Zhuomin and Sharma, Puru and Kang, Qingxuan and Jevdjic, Djordje and Deng, Junbo and Yang, Xingkun and Yu, Zhou and others}
}

@misc{gim_prompt_2024,
  title = {Prompt {{Cache}}: {{Modular Attention Reuse}} for {{Low-Latency Inference}}},
  shorttitle = {Prompt {{Cache}}},
  author = {Gim, In and Chen, Guojun and Lee, Seung-seob and Sarda, Nikhil and Khandelwal, Anurag and Zhong, Lin},
  year = 2024,
  month = apr,
  number = {arXiv:2311.04934},
  eprint = {2311.04934},
  primaryclass = {cs.CL},
  publisher = {arXiv},
  doi = {10.48550/arXiv.2311.04934},
  urldate = {2026-06-25},
  archiveprefix = {arXiv}
}

@article{grattafiori_llama_2024,
  title = {The Llama 3 Herd of Models},
  year = 2024,
  journal = {arXiv preprint arXiv:2407.21783},
  eprint = {2407.21783},
  archiveprefix = {arXiv},
  author = {Grattafiori, Aaron and Dubey, Abhimanyu and Jauhri, Abhinav and Pandey, Abhinav and Kadian, Abhishek and {Al-Dahle}, Ahmad and Letman, Aiesha and Mathur, Akhil and others}
}

@misc{gunasekar_textbooks_2023,
  title = {Textbooks {{Are All You Need}}},
  year = 2023,
  month = oct,
  number = {arXiv:2306.11644},
  eprint = {2306.11644},
  primaryclass = {cs.CL},
  publisher = {arXiv},
  doi = {10.48550/arXiv.2306.11644},
  urldate = {2026-06-25},
  archiveprefix = {arXiv},
  author = {Gunasekar, Suriya and Zhang, Yi and Aneja, Jyoti and Mendes, Caio C{\'e}sar Teodoro and Giorno, Allie Del and Gopi, Sivakanth and Javaheripi, Mojan and Kauffmann, Piero and others}
}

@misc{guo_deepseekcoder_2024,
  title = {{{DeepSeek-Coder}}: {{When}} the {{Large Language Model Meets Programming}} -- {{The Rise}} of {{Code Intelligence}}},
  shorttitle = {{{DeepSeek-Coder}}},
  year = 2024,
  month = jan,
  number = {arXiv:2401.14196},
  eprint = {2401.14196},
  primaryclass = {cs.SE},
  publisher = {arXiv},
  doi = {10.48550/arXiv.2401.14196},
  urldate = {2026-06-25},
  archiveprefix = {arXiv},
  author = {Guo, Daya and Zhu, Qihao and Yang, Dejian and Xie, Zhenda and Dong, Kai and Zhang, Wentao and Chen, Guanting and Bi, Xiao and others}
}

@inproceedings{hong_sola_2025,
  title = {{{SOLA}}: {{Optimizing SLO Attainment}} for {{Large Language Model Serving}} with {{State-Aware Scheduling}}},
  shorttitle = {{{SOLA}}},
  booktitle = {Eighth {{Conference}} on {{Machine Learning}} and {{Systems}}},
  year = 2025,
  month = may,
  urldate = {2026-01-29},
  langid = {english},
  author = {Hong, Ke and Li, Xiuhong and Chen, Lufang and Mao, Qiuli and Dai, Guohao and Ning, Xuefei and Yan, Shengen and Liang, Yun and others}
}

@misc{huggingface_text_2023,
  title = {Text {{Generation Inference}}},
  author = {{Hugging Face}},
  year = 2023
}

@inproceedings{imai_predicting_2024,
  title = {Predicting {{LLM Inference Latency}}: {{A Roofline-Driven ML Method}}},
  shorttitle = {Predicting {{LLM Inference Latency}}},
  booktitle = {Annual {{Conference}} on {{Neural Information Processing Systems}}},
  author = {Imai, Saki and Nakazawa, Rina and Amaral, Marcelo and Choochotkaew, Sunyanan and Chiba, Tatsuhiro},
  year = 2024,
  month = dec,
  urldate = {2026-01-29},
  copyright = {\copyright{} Copyright IBM Corp. 2021},
  langid = {american}
}

@misc{jia_dissecting_2018,
  title = {Dissecting the {{NVIDIA Volta GPU Architecture}} via {{Microbenchmarking}}},
  author = {Jia, Zhe and Maggioni, Marco and Staiger, Benjamin and Scarpazza, Daniele P.},
  year = 2018,
  month = apr,
  number = {arXiv:1804.06826},
  eprint = {1804.06826},
  primaryclass = {cs.DC},
  publisher = {arXiv},
  doi = {10.48550/arXiv.1804.06826},
  urldate = {2026-06-24},
  archiveprefix = {arXiv}
}

@misc{jiang_mistral_2023b,
  title = {Mistral {{7B}}},
  year = 2023,
  month = oct,
  number = {arXiv:2310.06825},
  eprint = {2310.06825},
  primaryclass = {cs.CL},
  publisher = {arXiv},
  doi = {10.48550/arXiv.2310.06825},
  urldate = {2026-06-25},
  archiveprefix = {arXiv},
  author = {Jiang, Albert Q. and Sablayrolles, Alexandre and Mensch, Arthur and Bamford, Chris and Chaplot, Devendra Singh and de las Casas, Diego and Bressand, Florian and Lengyel, Gianna and others}
}

@inproceedings{kwon_efficient_2023a,
  title = {Efficient Memory Management for Large Language Model Serving with Pagedattention},
  booktitle = {Proceedings of the 29th Symposium on Operating Systems Principles},
  year = 2023,
  pages = {611--626},
  author = {Kwon, Woosuk and Li, Zhuohan and Zhuang, Siyuan and Sheng, Ying and Zheng, Lianmin and Yu, Cody Hao and Gonzalez, Joseph and Zhang, Hao and others}
}

@inproceedings{lee_forecasting_2025,
  title = {Forecasting {{GPU Performance}} for {{Deep Learning Training}} and {{Inference}}},
  booktitle = {Proceedings of the 30th {{ACM International Conference}} on {{Architectural Support}} for {{Programming Languages}} and {{Operating Systems}}, {{Volume}} 1},
  author = {Lee, Seonho and Phanishayee, Amar and Mahajan, Divya},
  year = 2025,
  month = mar,
  eprint = {2407.13853},
  primaryclass = {cs},
  pages = {493--508},
  doi = {10.1145/3669940.3707265},
  urldate = {2026-01-29},
  archiveprefix = {arXiv}
}

@inproceedings{lee_infinigen_2024,
  title = {\textbraceleft{{InfiniGen}}\textbraceright : {{Efficient}} Generative Inference of Large Language Models with Dynamic \textbraceleft{{KV}}\textbraceright{} Cache Management},
  booktitle = {18th {{USENIX Symposium}} on {{Operating Systems Design}} and {{Implementation}} ({{OSDI}} 24)},
  author = {Lee, Wonbeom and Lee, Jungi and Seo, Junghwan and Sim, Jaewoong},
  year = 2024,
  pages = {155--172}
}

@article{lewis_retrievalaugmented_2020,
  title = {Retrieval-Augmented Generation for Knowledge-Intensive Nlp Tasks},
  year = 2020,
  journal = {Advances in neural information processing systems},
  volume = {33},
  pages = {9459--9474},
  author = {Lewis, Patrick and Perez, Ethan and Piktus, Aleksandra and Petroni, Fabio and Karpukhin, Vladimir and Goyal, Naman and K{\"u}ttler, Heinrich and Lewis, Mike and others}
}

@misc{lin_apex_2025,
  title = {{{APEX}}: {{An Extensible}} and {{Dynamism-Aware Simulator}} for {{Automated Parallel Execution}} in {{LLM Serving}}},
  shorttitle = {{{APEX}}},
  author = {Lin, Yi-Chien and Kwon, Woosuk and Pineda, Ronald and Paravecino, Fanny Nina},
  year = 2025,
  month = apr,
  number = {arXiv:2411.17651},
  eprint = {2411.17651},
  primaryclass = {cs},
  publisher = {arXiv},
  doi = {10.48550/arXiv.2411.17651},
  urldate = {2026-01-29},
  archiveprefix = {arXiv}
}

@article{mahmood_llmpowered_2023,
  title = {Llm-Powered Conversational Voice Assistants: {{Interaction}} Patterns, Opportunities, Challenges, and Design Guidelines},
  author = {Mahmood, Amama and Wang, Junxiang and Yao, Bingsheng and Wang, Dakuo and Huang, Chien-Ming},
  year = 2023,
  journal = {arXiv preprint arXiv:2309.13879},
  eprint = {2309.13879},
  archiveprefix = {arXiv}
}

@misc{microsoft_phi4mini_2025,
  title = {Phi-4-{{Mini Technical Report}}: {{Compact}} yet {{Powerful Multimodal Language Models}} via {{Mixture-of-LoRAs}}},
  shorttitle = {Phi-4-{{Mini Technical Report}}},
  year = 2025,
  month = mar,
  number = {arXiv:2503.01743},
  eprint = {2503.01743},
  primaryclass = {cs.CL},
  publisher = {arXiv},
  doi = {10.48550/arXiv.2503.01743},
  urldate = {2026-06-25},
  archiveprefix = {arXiv},
  author = {Microsoft and Abouelenin, Abdelrahman and Ashfaq, Atabak and Atkinson, Adam and Awadalla, Hany and Bach, Nguyen and Bao, Jianmin and Benhaim, Alon and others}
}

@misc{nvidiacorporation_nvidia_2024,
  title = {{{NVIDIA CUDA Toolkit}}, {{Version}} 12.x},
  author = {{NVIDIA Corporation}},
  year = 2024
}

@misc{nvidiacorporation_tensorrtllm_2023,
  title = {{{TensorRT-LLM}}: {{NVIDIA}}'s {{Inference Optimization Library}}},
  author = {{NVIDIA Corporation}},
  year = 2023
}

@article{patarasuk_bandwidth_2009,
  title = {Bandwidth Optimal All-Reduce Algorithms for Clusters of Workstations},
  author = {Patarasuk, Pitch and Yuan, Xin},
  year = 2009,
  month = feb,
  journal = {Journal of Parallel and Distributed Computing},
  volume = {69},
  number = {2},
  pages = {117--124},
  issn = {07437315},
  doi = {10.1016/j.jpdc.2008.09.002},
  urldate = {2026-06-24},
  langid = {english}
}

@inproceedings{patel_splitwise_2024,
  title = {Splitwise: {{Efficient Generative LLM Inference Using Phase Splitting}}},
  shorttitle = {Splitwise},
  booktitle = {2024 {{ACM}}/{{IEEE}} 51st {{Annual International Symposium}} on {{Computer Architecture}} ({{ISCA}})},
  author = {Patel, Pratyush and Choukse, Esha and Zhang, Chaojie and Shah, Aashaka and Goiri, {\'I}{\~n}igo and Maleki, Saeed and Bianchini, Ricardo},
  year = 2024,
  month = jun,
  pages = {118--132},
  publisher = {IEEE},
  address = {Buenos Aires, Argentina},
  doi = {10.1109/ISCA59077.2024.00019},
  urldate = {2026-06-24},
  copyright = {https://doi.org/10.15223/policy-029},
  isbn = {979-8-3503-2658-1}
}

@inproceedings{qi_paleo_2017,
  title = {Paleo: {{A Performance Model}} for {{Deep Neural Networks}}},
  shorttitle = {Paleo},
  booktitle = {International {{Conference}} on {{Learning Representations}}},
  author = {Qi, Hang and Sparks, Evan R. and Talwalkar, Ameet},
  year = 2017,
  month = feb,
  urldate = {2026-06-25},
  langid = {english}
}

@misc{qin_mooncake_2025,
  title = {Mooncake: {{A KVCache-centric Disaggregated Architecture}} for {{LLM Serving}}},
  shorttitle = {Mooncake},
  author = {Qin, Ruoyu and Li, Zheming and He, Weiran and Zhang, Mingxing and Wu, Yongwei and Zheng, Weimin and Xu, Xinran},
  year = 2025,
  month = sep,
  number = {arXiv:2407.00079},
  eprint = {2407.00079},
  primaryclass = {cs.DC},
  publisher = {arXiv},
  doi = {10.48550/arXiv.2407.00079},
  urldate = {2026-06-25},
  archiveprefix = {arXiv}
}

@misc{qwen_qwen25_2025,
  title = {Qwen2.5 {{Technical Report}}},
  year = 2025,
  month = jan,
  number = {arXiv:2412.15115},
  eprint = {2412.15115},
  primaryclass = {cs.CL},
  publisher = {arXiv},
  doi = {10.48550/arXiv.2412.15115},
  urldate = {2026-06-25},
  archiveprefix = {arXiv},
  author = {Qwen and Yang, An and Yang, Baosong and Zhang, Beichen and Hui, Binyuan and Zheng, Bo and Yu, Bowen and Li, Chengyuan and others}
}

@inproceedings{rashidi_astrasim_2020,
  title = {{{ASTRA-SIM}}: {{Enabling SW}}/{{HW Co-Design Exploration}} for {{Distributed DL Training Platforms}}},
  shorttitle = {{{ASTRA-SIM}}},
  booktitle = {2020 {{IEEE International Symposium}} on {{Performance Analysis}} of {{Systems}} and {{Software}} ({{ISPASS}})},
  author = {Rashidi, Saeed and Sridharan, Srinivas and Srinivasan, Sudarshan and Krishna, Tushar},
  year = 2020,
  month = aug,
  pages = {81--92},
  doi = {10.1109/ISPASS48437.2020.00018},
  urldate = {2026-01-29}
}

@article{reed_torch_2022,
  title = {Torch. Fx: {{Practical}} Program Capture and Transformation for Deep Learning in Python},
  author = {Reed, James and DeVito, Zachary and He, Horace and Ussery, Ansley and Ansel, Jason},
  year = 2022,
  journal = {Proceedings of Machine Learning and Systems},
  volume = {4},
  pages = {638--651}
}

@article{shah_flashattention3_2024,
  title = {Flashattention-3: {{Fast}} and Accurate Attention with Asynchrony and Low-Precision},
  author = {Shah, Jay and Bikshandi, Ganesh and Zhang, Ying and Thakkar, Vijay and Ramani, Pradeep and Dao, Tri},
  year = 2024,
  journal = {Advances in Neural Information Processing Systems},
  volume = {37},
  pages = {68658--68685}
}

@misc{shazeer_fast_2019,
  title = {Fast {{Transformer Decoding}}: {{One Write-Head}} Is {{All You Need}}},
  shorttitle = {Fast {{Transformer Decoding}}},
  author = {Shazeer, Noam},
  year = 2019,
  month = nov,
  number = {arXiv:1911.02150},
  eprint = {1911.02150},
  primaryclass = {cs.NE},
  publisher = {arXiv},
  doi = {10.48550/arXiv.1911.02150},
  urldate = {2026-06-25},
  archiveprefix = {arXiv}
}

@misc{shoeybi_megatronlm_2020a,
  title = {Megatron-{{LM}}: {{Training Multi-Billion Parameter Language Models Using Model Parallelism}}},
  shorttitle = {Megatron-{{LM}}},
  author = {Shoeybi, Mohammad and Patwary, Mostofa and Puri, Raul and LeGresley, Patrick and Casper, Jared and Catanzaro, Bryan},
  year = 2020,
  month = mar,
  number = {arXiv:1909.08053},
  eprint = {1909.08053},
  primaryclass = {cs.CL},
  publisher = {arXiv},
  doi = {10.48550/arXiv.1909.08053},
  urldate = {2026-06-25},
  archiveprefix = {arXiv}
}

@article{sun_dissecting_2023,
  title = {Dissecting {{Tensor Cores}} via {{Microbenchmarks}}: {{Latency}}, {{Throughput}} and {{Numeric Behaviors}}},
  shorttitle = {Dissecting {{Tensor Cores}} via {{Microbenchmarks}}},
  author = {Sun, Wei and Li, Ang and Geng, Tong and Stuijk, Sander and Corporaal, Henk},
  year = 2023,
  month = jan,
  journal = {IEEE Transactions on Parallel and Distributed Systems},
  volume = {34},
  number = {1},
  eprint = {2206.02874},
  primaryclass = {cs.AR},
  pages = {246--261},
  issn = {1045-9219, 1558-2183, 2161-9883},
  doi = {10.1109/TPDS.2022.3217824},
  urldate = {2026-06-25},
  archiveprefix = {arXiv}
}

@misc{team_gemma_2024,
  title = {Gemma 2: {{Improving Open Language Models}} at a {{Practical Size}}},
  shorttitle = {Gemma 2},
  year = 2024,
  month = oct,
  number = {arXiv:2408.00118},
  eprint = {2408.00118},
  primaryclass = {cs.CL},
  publisher = {arXiv},
  doi = {10.48550/arXiv.2408.00118},
  urldate = {2026-06-25},
  archiveprefix = {arXiv},
  author = {Team, Gemma and Riviere, Morgane and Pathak, Shreya and Sessa, Pier Giuseppe and Hardin, Cassidy and Bhupatiraju, Surya and Hussenot, L{\'e}onard and Mesnard, Thomas and others}
}

@misc{vaswani_attention_2023,
  title = {Attention {{Is All You Need}}},
  author = {Vaswani, Ashish and Shazeer, Noam and Parmar, Niki and Uszkoreit, Jakob and Jones, Llion and Gomez, Aidan N. and Kaiser, Lukasz and Polosukhin, Illia},
  year = 2023,
  month = aug,
  number = {arXiv:1706.03762},
  eprint = {1706.03762},
  primaryclass = {cs},
  publisher = {arXiv},
  doi = {10.48550/arXiv.1706.03762},
  urldate = {2025-03-05},
  archiveprefix = {arXiv}
}

@article{williams_roofline_2009,
  title = {Roofline: An Insightful Visual Performance Model for Multicore Architectures},
  shorttitle = {Roofline},
  author = {Williams, Samuel and Waterman, Andrew and Patterson, David},
  year = 2009,
  month = apr,
  journal = {Commun. ACM},
  volume = {52},
  number = {4},
  pages = {65--76},
  issn = {0001-0782},
  doi = {10.1145/1498765.1498785},
  urldate = {2026-01-29}
}

@inproceedings{won_astrasim20_2023,
  title = {{{ASTRA-sim2}}.0: {{Modeling Hierarchical Networks}} and {{Disaggregated Systems}} for {{Large-model Training}} at {{Scale}}},
  shorttitle = {{{ASTRA-sim2}}.0},
  booktitle = {2023 {{IEEE International Symposium}} on {{Performance Analysis}} of {{Systems}} and {{Software}} ({{ISPASS}})},
  author = {Won, William and Heo, Taekyung and Rashidi, Saeed and Sridharan, Srinivas and Srinivasan, Sudarshan and Krishna, Tushar},
  year = 2023,
  month = apr,
  eprint = {2303.14006},
  primaryclass = {cs},
  pages = {283--294},
  doi = {10.1109/ISPASS57527.2023.00035},
  urldate = {2026-01-29},
  archiveprefix = {arXiv}
}

@misc{wu_tokensim_2025,
  title = {{{TokenSim}}: {{Enabling Hardware}} and {{Software Exploration}} for {{Large Language Model Inference Systems}}},
  shorttitle = {{{TokenSim}}},
  year = 2025,
  month = mar,
  number = {arXiv:2503.08415},
  eprint = {2503.08415},
  primaryclass = {cs},
  publisher = {arXiv},
  doi = {10.48550/arXiv.2503.08415},
  urldate = {2026-01-29},
  archiveprefix = {arXiv},
  author = {Wu, Feiyang and Bian, Zhuohang and Duan, Guoyang and Xu, Tianle and Wu, Junchi and Ma, Teng and Yao, Yongqiang and Gong, Ruihao and others}
}

@misc{xu_aiconfigurator_2026,
  title = {{{AIConfigurator}}: {{Lightning-Fast Configuration Optimization}} for {{Multi-Framework LLM Serving}}},
  shorttitle = {{{AIConfigurator}}},
  year = 2026,
  month = jan,
  number = {arXiv:2601.06288},
  eprint = {2601.06288},
  primaryclass = {cs.LG},
  publisher = {arXiv},
  doi = {10.48550/arXiv.2601.06288},
  urldate = {2026-06-22},
  archiveprefix = {arXiv},
  author = {Xu, Tianhao and Liu, Yiming and Lu, Xianglong and Zhao, Yijia and Zhou, Xuting and Feng, Aichen and Chen, Yiyi and Shen, Yi and others}
}

@misc{yang_qwen3_2025b,
  title = {Qwen3 {{Technical Report}}},
  year = 2025,
  month = may,
  number = {arXiv:2505.09388},
  eprint = {2505.09388},
  primaryclass = {cs.CL},
  publisher = {arXiv},
  doi = {10.48550/arXiv.2505.09388},
  urldate = {2026-06-24},
  archiveprefix = {arXiv},
  author = {Yang, An and Li, Anfeng and Yang, Baosong and Zhang, Beichen and Hui, Binyuan and Zheng, Bo and Yu, Bowen and Gao, Chang and others}
}

@inproceedings{yu_habitat_2021,
  title = {Habitat: {{A Runtime-Based Computational Performance Predictor}} for {{Deep Neural Network Training}}},
  shorttitle = {Habitat},
  booktitle = {2021 {{USENIX Annual Technical Conference}} ({{USENIX ATC}} 21)},
  author = {Yu, Geoffrey X. and Gao, Yubo and Golikov, Pavel and Pekhimenko, Gennady},
  year = 2021,
  pages = {503--521},
  urldate = {2026-06-25},
  isbn = {978-1-939133-23-6},
  langid = {english}
}

@inproceedings{zhang_nnmeter_2021,
  title = {Nn-{{Meter}}: Towards Accurate Latency Prediction of Deep-Learning Model Inference on Diverse Edge Devices},
  shorttitle = {Nn-{{Meter}}},
  booktitle = {Proceedings of the 19th {{Annual International Conference}} on {{Mobile Systems}}, {{Applications}}, and {{Services}}},
  author = {Zhang, Li Lyna and Han, Shihao and Wei, Jianyu and Zheng, Ningxin and Cao, Ting and Yang, Yuqing and Liu, Yunxin},
  year = 2021,
  month = jun,
  pages = {81--93},
  publisher = {ACM},
  address = {Virtual Event Wisconsin},
  doi = {10.1145/3458864.3467882},
  urldate = {2026-06-25},
  isbn = {978-1-4503-8443-8},
  langid = {english}
}

@misc{zheng_sglang_2024,
  title = {{{SGLang}}: {{Efficient Execution}} of {{Structured Language Model Programs}}},
  shorttitle = {{{SGLang}}},
  year = 2024,
  month = jun,
  number = {arXiv:2312.07104},
  eprint = {2312.07104},
  primaryclass = {cs.AI},
  publisher = {arXiv},
  doi = {10.48550/arXiv.2312.07104},
  urldate = {2026-06-25},
  archiveprefix = {arXiv},
  author = {Zheng, Lianmin and Yin, Liangsheng and Xie, Zhiqiang and Sun, Chuyue and Huang, Jeff and Yu, Cody Hao and Cao, Shiyi and Kozyrakis, Christos and others}
}

@misc{zhong_distserve_2024,
  title = {{{DistServe}}: {{Disaggregating Prefill}} and {{Decoding}} for {{Goodput-optimized Large Language Model Serving}}},
  shorttitle = {{{DistServe}}},
  author = {Zhong, Yinmin and Liu, Shengyu and Chen, Junda and Hu, Jianbo and Zhu, Yibo and Liu, Xuanzhe and Jin, Xin and Zhang, Hao},
  year = 2024,
  month = jun,
  number = {arXiv:2401.09670},
  eprint = {2401.09670},
  primaryclass = {cs.DC},
  publisher = {arXiv},
  doi = {10.48550/arXiv.2401.09670},
  urldate = {2026-06-25},
  archiveprefix = {arXiv}
}



\appendix

\onecolumn

\section{Experimental Setup Details}
\label{app:setup}

\subsection{Software Version}

We use the NGC PyTorch container image (26.04-py3) with vLLM v0.19.0 and CUDA 13, which is portable across platforms.

\subsection{Evaluation Metric}

\paragraph{Mean Absolute Percentage Error: MAPE}

\begin{equation}
\mathrm{MAPE} = \frac{100\%}{n} \sum_{i=1}^{n} \left| \frac{\hat{y}_i - y_i}{y_i} \right|
\label{eq:mape}
\end{equation}

\paragraph{Per-Request Absolute Percent Error}

\begin{equation}
e_i = 100\% \cdot \frac{\lvert \hat{y}_i - y_i \rvert}{y_i}
\label{eq:abs-pct-err}
\end{equation}

\paragraph{Error Percentiles: p50, p90, p99}

\begin{equation}
\mathrm{P}_p = e_{(k)}, \qquad k = \left\lceil \tfrac{p}{100}\, n \right\rceil,
\qquad e_{(1)} \le e_{(2)} \le \dots \le e_{(n)}
\label{eq:percentile}
\end{equation}

\subsection{GPUs and EC2 Instances}

\begin{table}[ht]
\centering \small \setlength{\tabcolsep}{3.5pt}
\caption{GPUs used in our analysis and evaluation. They are deployed on AWS and the Boston University Shared Computing Clusters (BU SCC).
BF16 throughput is dense. Memory bandwidth is per GPU. Peaks vary by SKU/clocks.}

\label{tab:gpu_specs}
\resizebox{\columnwidth}{!}{%
\begin{tabular}{l l c c c c c c l}
\toprule
\textbf{GPU} & \textbf{Instance} & \textbf{GPUs} & \textbf{Arch} & \textbf{Year} &
\textbf{Mem / GPU} & \textbf{BW / GPU} & \textbf{BF16 TC} & \textbf{Mem type \& interconnect} \\
\midrule
A100-40GB      & p4d.24xlarge   & 8 & Ampere       & 2020 & 40\,GB  & $\sim$1.56\,TB/s & $\sim$312\,TFLOPS  & HBM2e; NVLink \\
A100-80GB-SXM4 & p4de.24xlarge  & 8 & Ampere       & 2020 & 80\,GB  & $\sim$2.04\,TB/s & $\sim$312\,TFLOPS  & HBM2e; NVLink \\
A10G           & g5.48xlarge    & 8 & Ampere       & 2021 & 24\,GB  & $\sim$0.60\,TB/s & $\sim$70\,TFLOPS   & GDDR6; PCIe \\
H100-80GB      & p5.48xlarge    & 8 & Hopper       & 2022 & 80\,GB  & $\sim$3.35\,TB/s & $\sim$989\,TFLOPS  & HBM3;  NVLink \\
H200-141GB     & p5e.48xlarge   & 8 & Hopper       & 2024 & 141\,GB & $\sim$4.8\,TB/s  & $\sim$989\,TFLOPS  & HBM3e; NVLink \\
L4             & g6.48xlarge    & 8 & Ada Lovelace & 2023 & 24\,GB  & $\sim$0.30\,TB/s & $\sim$121\,TFLOPS  & GDDR6; PCIe \\
GB200          & p6e-gb200.36xlarge & 4 & Blackwell & 2024 & 192\,GB & $\sim$8.0\,TB/s  & $\sim$2250\,TFLOPS & HBM3e; NVLink-5 \\
\midrule
A40            & BU SCC & 4 & Ampere       & 2020 & 48\,GB  & $\sim$0.70\,TB/s & $\sim$150\,TFLOPS  & GDDR6; PCIe \\
A100-80GB-PCIe & BU SCC & 4 & Ampere       & 2021 & 80\,GB  & $\sim$1.94\,TB/s & $\sim$312\,TFLOPS  & HBM2e; PCIe \\
L40S           & BU SCC & 4 & Ada Lovelace & 2023 & 48\,GB  & $\sim$0.86\,TB/s & $\sim$362\,TFLOPS  & GDDR6; PCIe \\
H200-NVL       & BU SCC & 4 & Hopper       & 2024 & 141\,GB & $\sim$4.8\,TB/s  & $\sim$836\,TFLOPS  & HBM3e; NVLink \\
\midrule
GB10 (RTX Spark)   & Local Deployment & 1 & Blackwell    & 2025 & 128\,GB & $\sim$0.27\,TB/s & $\sim$125\,TFLOPS  & LPDDR5X; NVLink-C2C \\
RTX 5090       & Local  Deployment    & 1 & Blackwell    & 2025 & 32\,GB  & $\sim$1.79\,TB/s & $\sim$419\,TFLOPS  & GDDR7; PCIe \\
\bottomrule
\end{tabular}}
\end{table}

\clearpage

\subsection{Models and Variants}

\begin{table}[ht]\centering\small
\caption{Model families and variants used across profiling, analysis, and evaluation.}
\label{tab:models}
\begin{tabular}{l l l c}
\toprule
\textbf{Family} & \textbf{Developer} & \textbf{Variants (parameters)} & \textbf{TP} \\
\midrule
Llama~\cite{grattafiori_llama_2024}    & Meta        & 3.2-1B, 3.2-3B, 3.1-8B, 3.3-70B            & 1,2,4 \\
Qwen~\cite{qwen_qwen25_2025, yang_qwen3_2025b}     & Alibaba     & 2.5-0.5B, 3-4B, 3-8B, 3-14B, 3-32B         & 1,2,4 \\
DeepSeek~\cite{bi_deepseek_2024, guo_deepseekcoder_2024} & DeepSeek-AI & LLM-7B, Coder-33B, LLM-67B                 & 1,2,4 \\
Phi~\cite{gunasekar_textbooks_2023, microsoft_phi4mini_2025, abdin_phi4_2024}      & Microsoft   & Phi-2 (2.7B), Phi-4-mini (3.8B), Phi-4 (14B) & 1 \\
Gemma~\cite{team_gemma_2024}    & Google      & Gemma-2-9B                                 & 1 \\
Mistral~\cite{jiang_mistral_2023b}  & Mistral AI  & 7B-Instruct-v0.3                           & 1 \\
\bottomrule
\end{tabular}
\end{table}

\clearpage

\section{Limitations}
\label{sec:limitations}

\subsection{Cross-Generation Kernel Prediction Challenge}

Our predictor factors each kernel's latency using the efficiency factor, $\eta$. This is a learned, dimensionless efficiency residual.
Cross-generation prediction (Tier~A) therefore rests on a single
assumption: that $\eta$, fit on a set of training GPUs, transfers to an unseen
target as claimed in \cite{lee_forecasting_2025}. In practice, this assumption holds more weakly than prior work suggests.

\begin{figure*}[ht]
  \centering
  \includegraphics[width=\textwidth]{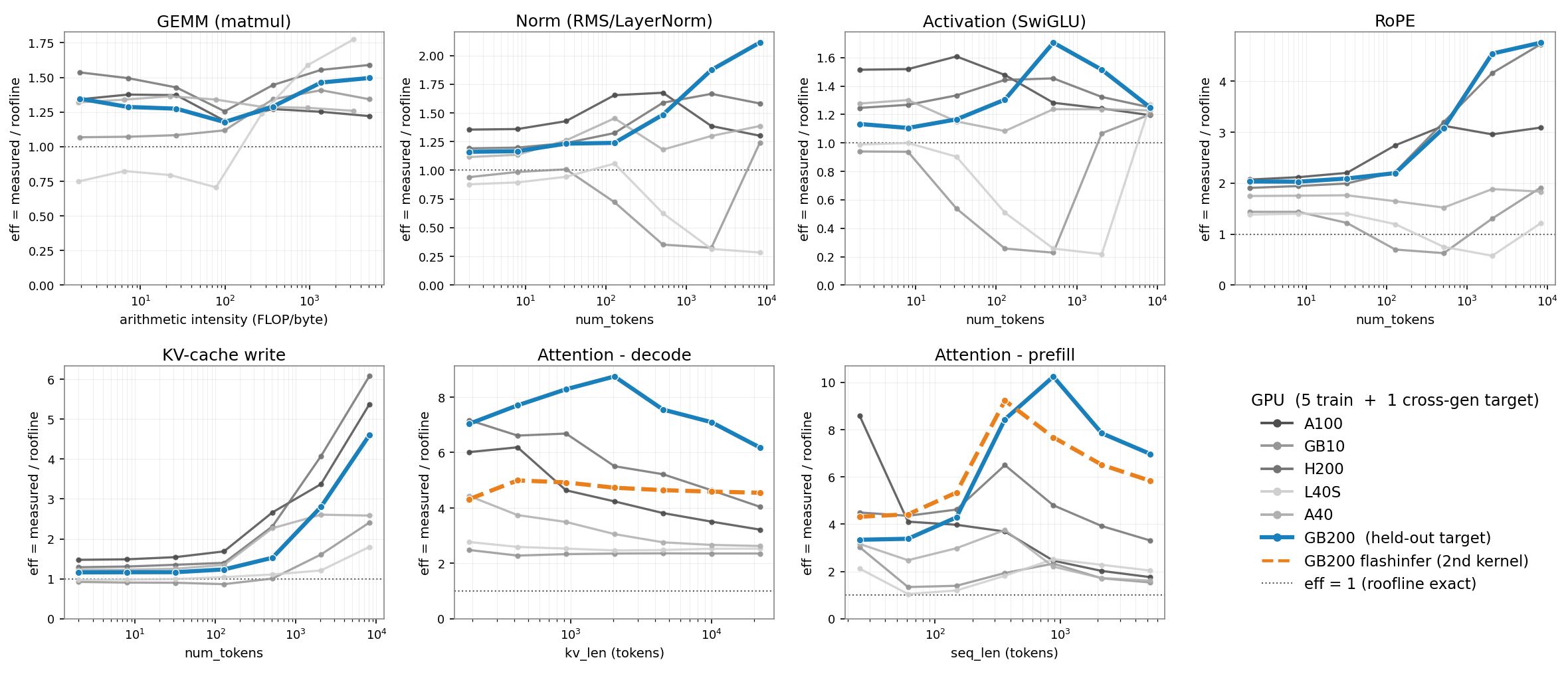}
  \caption{\textbf{The cross-generation test.} Per-kernel
  $\eta=\text{measured}/\text{roofline}$ versus each family's dominant shape
  parameter, across six GPUs. The five analysis devices are gray; GB200 (blue) is
  the held-out target. Where GB200's curve lies within the analysis envelope (GEMM), $\eta$
  interpolates; where it lies apart (attention, KV-cache, RoPE), $\eta$ is
  device-specific and cannot be recovered from the other devices. Dashed:
  GB200's second dispatched attention kernel.}
  \label{fig:crossgen-grouped}
\end{figure*}

Figure~\ref{fig:crossgen-grouped} performs the experiment directly: we hold out
GB200 and ask whether the analysis devices bracket its efficiency curve. For
compute-bound GEMM, the target lies inside the analysis envelope, so
interpolating $\eta$ recovers it. For the memory-bound families and especially
attention, the GB200 curve sits well above the cluster: its $\eta$ is set by an
achieved-bandwidth ceiling that does \emph{not} scale with peak bandwidth, and
for attention by the specific kernel the engine dispatches (the dashed line is a
second GB200 attention kernel with a markedly different curve). Neither is
observable from the analysis devices.

\begin{figure*}[ht]
  \centering
  \includegraphics[width=0.7\columnwidth]{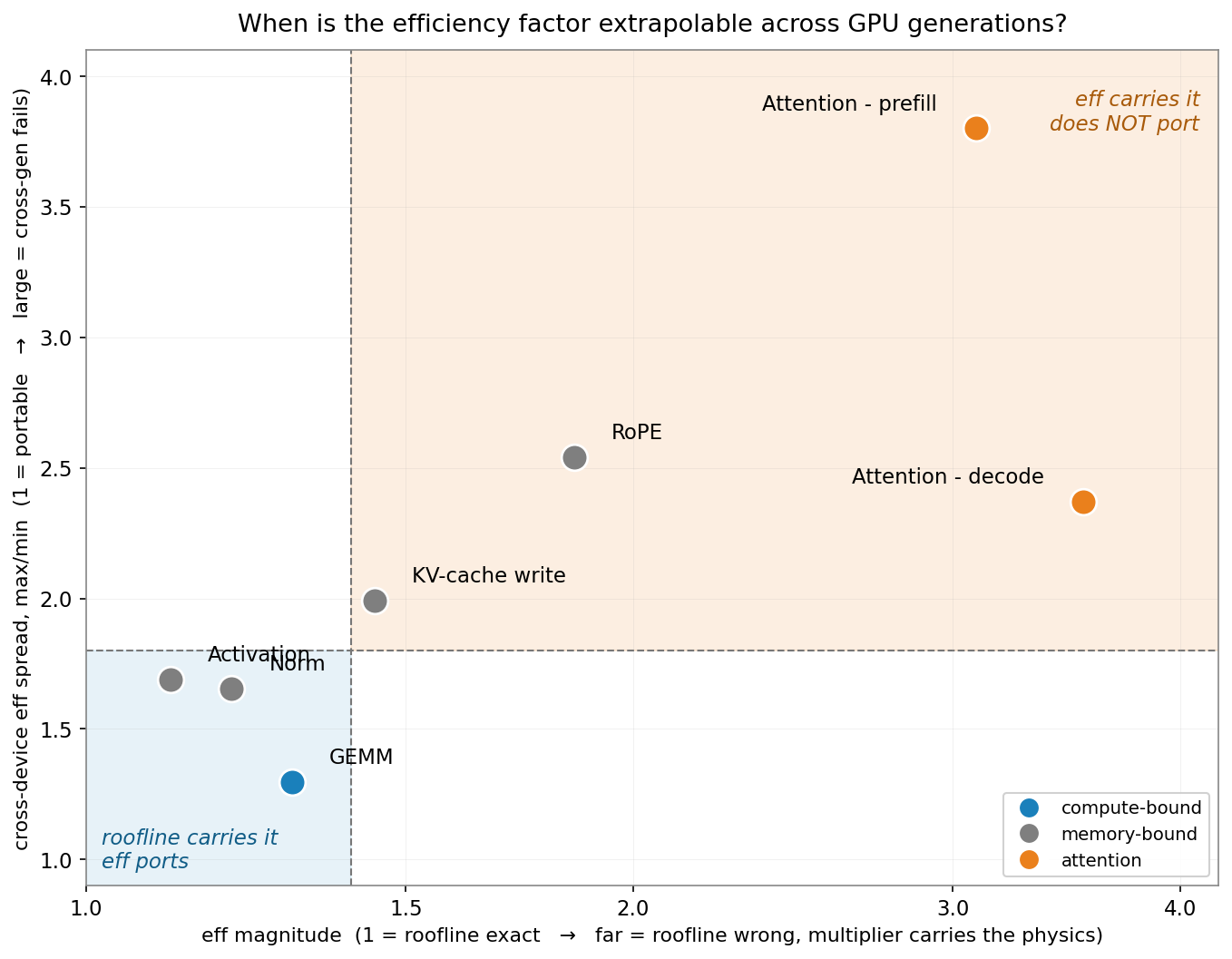}
  \caption{\textbf{When $\eta$ extrapolates.} Each family placed by $\eta$
  magnitude (geomean over devices; $1=$ roofline exact) and cross-device spread
  (max/min; $1=$ portable). The safe region (blue) holds families where the
  roofline is accurate and $\eta$ ports; the danger region (orange) holds those
  where $\eta$ both carries the prediction and varies across devices.}
  \label{fig:crossgen-zones}
\end{figure*}

Figure~\ref{fig:crossgen-zones} reduces each family to two numbers: the
magnitude of $\eta$ (how far the multiplier departs from $1$, i.e.\ how much of
the latency the roofline fails to explain) and its cross-device spread (its
portability). The families separate cleanly. GEMM occupies the safe corner
($\eta\!\approx\!1.3$, spread $1.3\times$): the roofline does the work and the
residual ports. Both attention variants occupy the opposite corner
($\eta\!\approx\!3$--$3.5$, spread up to $3.8\times$): the multiplier carries the
physics \emph{and} is device-specific. The diagonal between them is the core
limitation, \emph{the families for which the learned efficiency factor matters
most are precisely those for which it is least transferable.}

Two failure modes follow. When $\eta$ is far from $1$, the roofline acts as a
normalizer rather than a model, and accuracy depends entirely on the residual;
when $\eta$ varies several-fold across devices, no dimensionless feature
combination recovers it for an unseen generation (the best linear combination of
all shape features still leaves a large device-specific offset on these
families). This is tolerable for GEMM and the lighter memory-bound kernels, but
not for attention, where closing the gap requires at least one serving-matched
measurement on each new architecture rather than pure extrapolation.

\clearpage

\section{Behavior at saturation}
Another limitation is that the TTFT accuracy degrades as the offered rate approaches sustained capacity. Normalizing load as $\rho=\lambda/\mu$ (offered throughput over measured capacity), our TTFT error rises from $15\%$ to $22\%$ for $\rho\!\lesssim\!0.9$ but reaches $74\%$ at $\rho\!\approx\!1.2$, while TPOT and throughput errors stay $\le\!13\%$ and $\le\!7\%$. 

This is because at high utilization, the request queue builds up. This inflates $13\%$ per-step errors into a $74\%$ TTFT error that no service-rate-based method can resolve. In practice this regime rarely matters: latency-SLO-oriented deployments typically operate at $\rho\!\approx\!0.6$--$0.8$. Because our simulator is accurate below saturation, it remains useful as a practical reference.

\begin{figure}[ht]\centering
\includegraphics[width=0.5\columnwidth]{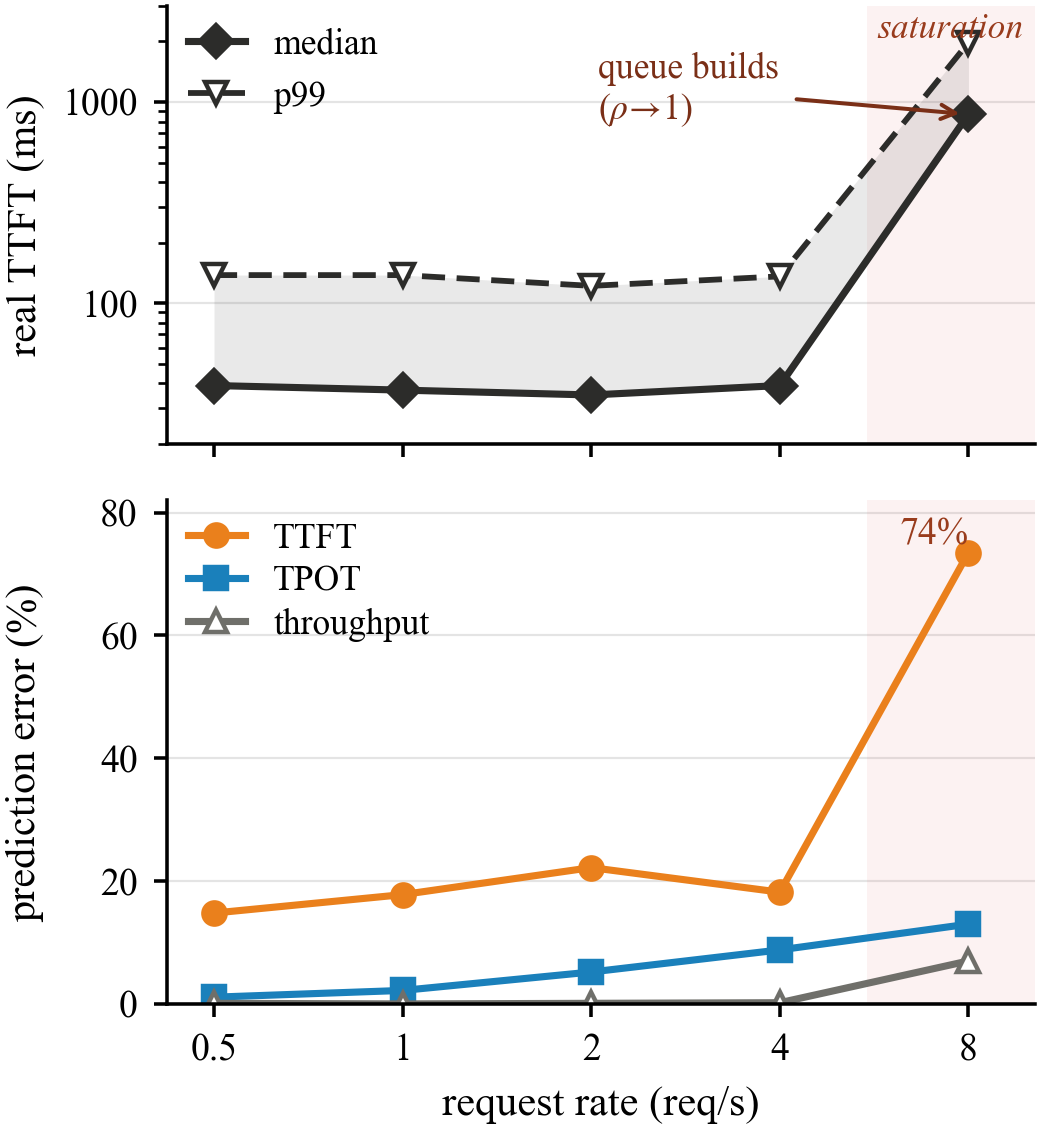}
\caption{Prediction error vs.\ offered load (One run of Qwen3, TP1, GB200). }
\label{fig:saturation}
\end{figure}




\end{document}